\documentclass{emulateapj}
\renewcommand{\*}{$^*$}

\newcommand{\etal}{et al.}

\newcommand{\uv}{\mbox{$u$-$v$}}

\newcommand{\kms}{\mbox{km s$^{-1}$}}
\newcommand{\kmsMpc}{\mbox{km s$^{-1}$ Mpc$^{-1}$}}
\newcommand{\muas}{\mbox{$\mu$as}}

\newcommand{\thout}{\mbox{$\theta_{\rm o}$}}
\newcommand{\thin}{\mbox{$\theta_{\rm i}$}}
\newcommand{\thc}{\mbox{$\theta_{\rm c}$}}
\newcommand{\thx}{\mbox{$\theta_{\rm x}$}}
\newcommand{\dotthout}{\mbox{$\dot \theta_{\rm o}$}}
\newcommand{\dotthin}{\mbox{$\dot \theta_{\rm i}$}}
\newcommand{\dotthc}{\mbox{$\dot \theta_{\rm c}$}}
\newcommand{\dotthx}{\mbox{$\dot \theta_{\rm x}$}}

\shortauthors{Bartel  et al.}
\shorttitle{SN~1993J VLBI IV}

\begin{document}
\title{SN~1993J VLBI (IV): A Geometric 
Determination of the Distance to M81 with the Expanding Shock Front Method}

\author{N. Bartel and M. F. Bietenholz}
\affil{Department of Physics and Astronomy, York University, Toronto, ON M3J~1P3, Canada}

\author{M. P. Rupen}
\affil{National Radio Astronomy Observatory, Socorro, NM 87801, USA}

\author{and V. V. Dwarkadas}
\affil{Department of Astronomy and Astrophysics, University of Chicago, Chicago, IL 60637, USA}
\shortauthors{Bartel \etal}
\shorttitle{Distance to M81}

\author{(Accepted for publication in ApJ, tentatively scheduled for the Oct. 20, 2007, 
v668n 2 issue)}

\begin{abstract}

We compare the angular expansion velocities, determined with VLBI, with
the linear expansion velocities measured from optical spectra for
supernova 1993J in the galaxy M81, over the period from 7~d to
$\sim$9~yr after shock breakout.  The high degree of isotropy of the
radio shell's expansion, within 5.5\%, with the projection of the
radio shell being circular within even 1.4\%, and the consistency of
the radio shell thickness with predictions from hydrodynamic
simulations and analytical computations argue strongly in favor of the
radio-emitting shell being bound by the forward and reverse shocks.
The absorption and emission of hydrogen at the highest velocity most
likely also arise between the forward and reverse shocks, specifically
in the shocked ejecta behind the contact surface which is close to the
reverse shock, and possibly in Rayleigh-Taylor fingers that extend
beyond it, farther out into the radio shell but not beyond the forward
shock. The radio shell and the H$\alpha$ absorbing and emitting gas
are on average similarly decelerated, but the latter slightly less so
than the former several years after shock breakout.  This may indicate
developing Rayleigh-Taylor fingers, extending progressively further
into the shocked circumstellar medium.  We estimate the distance to
SN~1993J using the Expanding Shock Front Method (ESM).  We find the
best distance estimate is obtained by fitting the angular velocity of
a point halfway between the contact surface and outer shock front to
the maximum observed hydrogen gas velocity.  We obtain a direct,
geometric, distance estimate for M81 of $D=3.96\pm0.05\pm0.29$~Mpc with
statistical and systematic error contributions, respectively,
corresponding to a total standard error of $\pm0.29$~Mpc.  The upper
limit of 4.25 Mpc corresponds to the hydrogen gas with the highest
observed velocity reaching no farther out than the contact surface a
few days after shock breakout.  The lower limit of 3.67 Mpc
corresponds to this hydrogen gas reaching as far out as the forward
shock for the whole period, which would mean that Rayleigh-Taylor
fingers have grown to the forward shock already a few days after shock
breakout. Our distance estimate is $9\pm13$\% larger than that of
$3.63\pm0.34$~Mpc from the HST Key Project, which is near our lower limit
but within the errors.

\end{abstract}


\section{INTRODUCTION}
Supernova 1993J was discovered in a spiral arm of M81 south south-west
of the galaxy's center by Garcia (Ripero \& Garcia, 1993) on 28 March
1993, shortly after shock breakout at $\sim$ 0 UT (Wheeler \etal\
1993) on the same day ($t=0$~d). It subsequently became the optically
brightest supernova in the northern hemisphere since SN 1954A and
one of the brightest radio supernovae ever detected. It is also one of
the closest extragalactic supernovae ever observed and is second only
to SN 1987A as a subject of intense observational and theoretical
supernova studies. This combination allows for sensitive spectroscopic
and spectropolarimetric observations and for exceptionally detailed
VLBI imaging, the latter giving us the highest relative resolution
ever obtained for any radio supernova (Figure~\ref{f1m81sn}).

Combining the radial velocities of the ejecta gas obtained from the
optical lines with the transverse {\em angular} velocities of the
radio shell obtained from VLBI measurements yields a direct estimate
of the distance to the supernova and its host galaxy. This method,
called the Expanding Shock Front Method (ESM), was used to determine
the distance to SN 1979C in M100 in the Virgo cluster (Bartel 1985;
Bartel \etal\ 1985; Bartel \& Bietenholz 2003, 2005). However the
relatively large distance to Virgo has made it hard to resolve the
supernova in sufficient detail, and an accurate distance estimate for
Virgo is still pending.

In contrast, M81, the host galaxy of SN~1993J, is much closer.  Its
distance was recently determined via HST observations of Cepheids to
be $3.63\pm0.34$~Mpc (Freedman \etal\ 1994, see also Freedman \etal\ 2001, HST Key Project) 
and 3.93$\pm$0.26~Mpc (Huterer, Sasselov, \& Schechter, 1995).
Here we report on a detailed comparison between the angular velocities
obtained from the VLBI measurements of the radio shell radius and the
radial velocities obtained from the Doppler shifts of H$\alpha$,
H$\beta$, He I, O[III], and Na I optical lines and determine the
distance to SN~1993J and its host galaxy M81 with ESM.

This paper is the fourth in a series, presenting the results from our
VLBI campaign on this supernova (see Bartel et~al. 1994 for early
results and Bartel et~al. 2000 for an introduction of this series of
results, see Marcaide et~al. 1997 for parallel observations). In the
first paper (Bietenholz, Bartel, \& Rupen 2001; Paper~I), we located
the explosion center with respect to the nuclear radio source of M81,
thus defining a stable reference point for our images.  We also
determined, using model-fitting, the motion of the geometric center of
SN 1993J, and comment on the high degree of circular symmetry shown by
SN 1993J\@.  In the second paper (Bartel et al.\ 2002; Paper~II), we
determined the expansion speed of SN 1993J and measured its
deceleration as a function of time.  In the third paper, we presented
a complete series of VLBI images of SN 1993J at 8.4 and 5.0~GHz, along
with our latest image at 1.7~GHz (Bietenholz, Bartel, \& Rupen, 2003;
Paper~III). In this fourth paper we use the main results of each of
the previous papers, compare them with optical spectroscopic
observations by others, include specific predictions from hydrodynamic
simulations of this supernova for comparison with our observations,
and determine directly the distance to M81.

In \S~\ref{sobsdat}, we summarize our VLBI observations and data
reduction.  In \S~\ref{snmodel} we describe the generic supernova
shell model with its radio and optical emission regions and 
discuss the origin of the radio and optical
emission and the relation of the forward and reverse shock to the
measured boundaries of the radio emitting shell. In \S~\ref{exp} we
elaborate on the expansion of the supernova as measured in the radio
and optical wavelength range.  Then we determine the distance to
SN~1993J and its host galaxy in \S~\ref{dist}, discuss our results in
\S~\ref{disc} and finally give our conclusions in \S~\ref{conc}.

\section {OBSERVATIONS AND DATA REDUCTION}
\label{sobsdat}

The observations were described in Papers~I and II\@.  To summarize, we
observed SN~1993J at 34 epochs between 1993 and 2001 from $t=30$ to
$\sim3000$~d.  At the earliest epoch, at $t=30$~d, we observed only at
22.2 GHz. At later epochs up to the $33^{\rm rd}$ epoch at $t=2996$~d,
we observed mostly at 8.4 GHz and often also at 5.0 GHz. In addition,
we also observed at some early epochs at 14.8 GHz and throughout the
full eight years sporadically also at 2.3 and 1.7 GHz.  We used a
global array of between 9 and 18 telescopes with a total time of 9 to
18 hours for each session.  We have continued observing SN~1993J at
approximately one-year intervals and will report on these observations
in future papers.

The data were recorded with either the MK~III or the VLBA/MKIV VLBI
systems, and correlated with the NRAO VLBA processor in Socorro, New
Mexico, USA\@. We refer the interested reader to Paper~II where details
of the observing sessions are tabulated.  The analysis was carried out
using NRAO's Astronomical Image Processing System (AIPS).

Most of our observations were phase-referenced to M81\*, the radio
source in the center of the galaxy M81.  Being very compact
(Bietenholz et~al.\ 1996), M81\* is an excellent calibrator for
phase-referenced mapping.  Our images are among those with the lowest
background noise level currently obtained with VLBI\@. Bietenholz,
Bartel, \& Rupen (2000) located a fixed point in the variable
brightness distribution of M81\* that also has an inverted radio
spectrum (Bietenholz, Bartel, \& Rupen 2004a).  We identified this
point with the purported supermassive black hole in the center of the
galaxy. Using this point as a reference point allowed us to locate the
explosion center of SN~1993J on each of our supernova images with high
precision.

\begin{figure}
\plotone{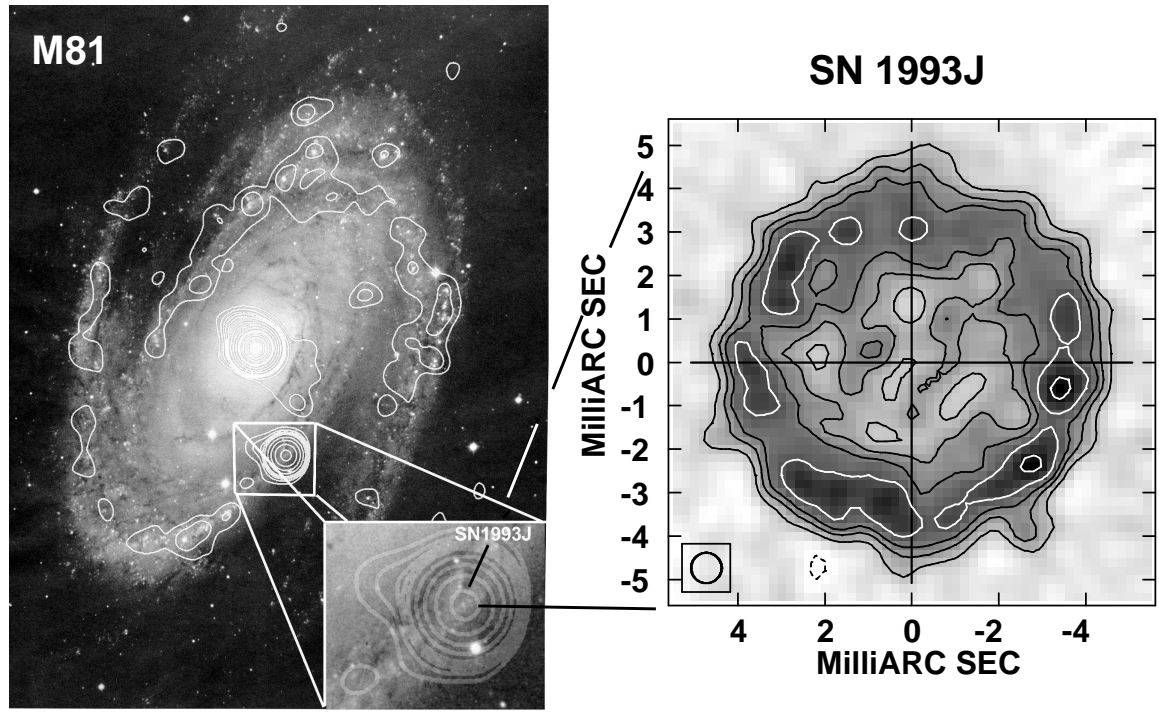}
\figcaption{Left panel: An optical image of the spiral galaxy M81 
(Sandage 1961) with a VLA contour image overlaid and with SN~1993J
prominent as the brightest radio source after the nucleus of the
galaxy.  Right panel: A composite VLBI image of SN~1993J at 8.4 GHz
from three epochs at $t = 2080$, 2525 and 2787~d (1998 December to
2000 November) all corrected for position shifts, scaled in flux
density and radius to the values of the 2000 November data, and then
combined (adopted from Paper III). The beam (FWHM, 0.70~mas) is shown
at lower left.  The contours are drawn at $-16$, 16, 32, 45.3, 64 and
90\% of the peak brightness, and the background rms was 5.3\% of the
peak brightness. North is up and east to the left.
\label{f1m81sn}}
\end{figure}

In Figure~\ref{f1m81sn} we show the galaxy M81 at optical
wavelengths. Overlayed is an image of the galaxy at radio wavelengths,
with SN~1993J clearly visible in a southern spiral arm. On the right
side of the figure we show the VLBI image of the evolved radio shell
which has the highest relative resolution obtained so far (Paper~III).

In addition to imaging, we used model-fitting. We fit, by weighted
least-squares, the two-dimensional projection of a three-dimensional
spherical shell of uniform volume emissivity to the calibrated
\uv~data consistently for all epochs. The ratio of the outer to inner
angular radius was fixed at \thout/\thin = 1.25. From this fit, we
estimated the shell's center coordinates, $x,~y$, (see Paper~I), and
\thout\ (see Paper~II).  The shell's center coordinates were found to
be equal, within an rms of 64~\muas, to those of the explosion center
of SN 1993J\@. In the image in Figure~\ref{f1m81sn} this center is at
the origin. In addition, for relatively late epochs, we also
determined the shell thickness by freeing the ratio of the radii and
estimating \thout\ and \thin\ independently.

\section{THE GENERIC SUPERNOVA SHELL MODEL AND ITS RELEVANCE FOR ESM}
\label{snmodel}

\begin{figure}
\includegraphics[width=1.0\textwidth, trim=0 4.5in 2in 0in,clip]{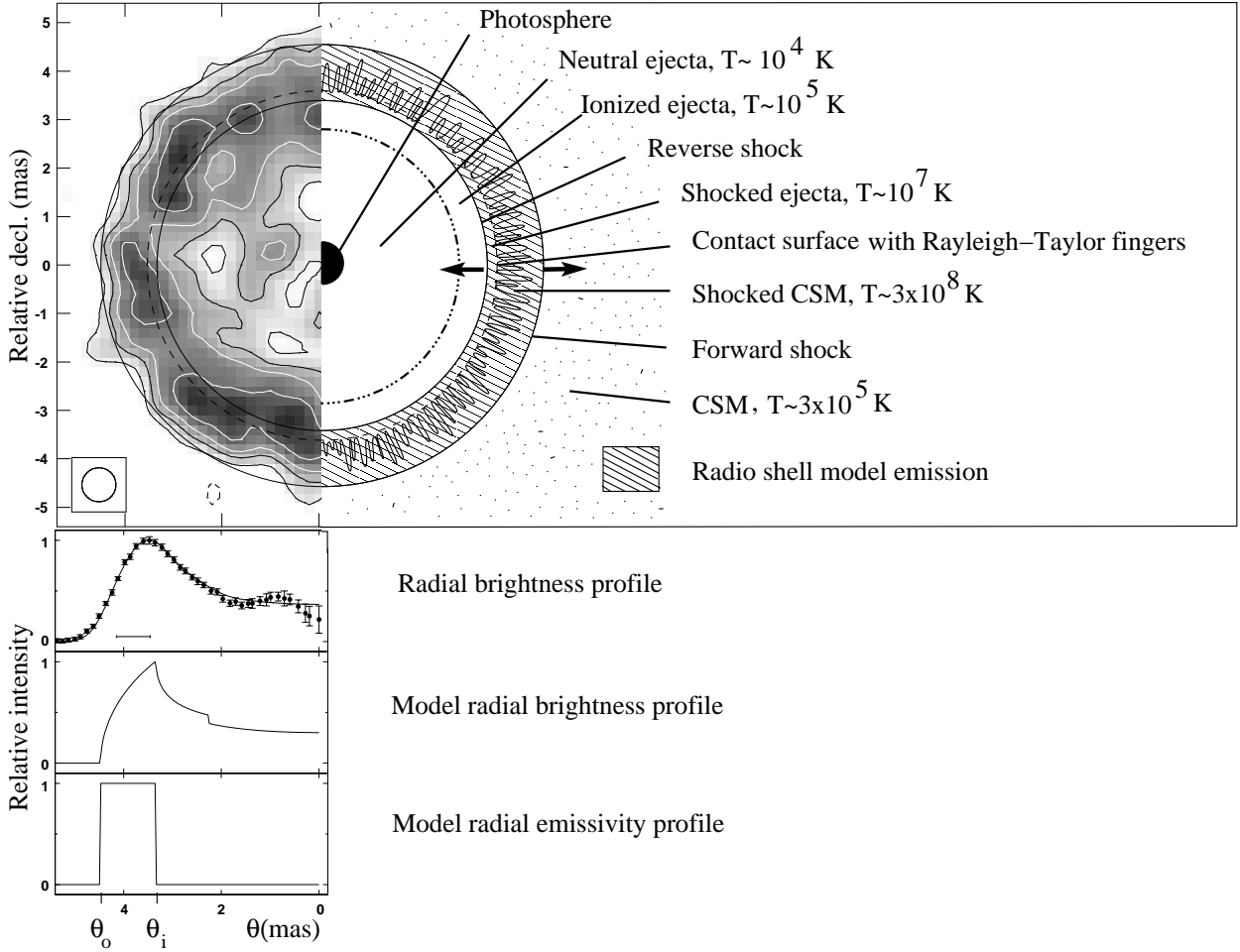}
\figcaption{\scriptsize Top panel: The left part of the composite VLBI image of
SN~1993J from Figure~\ref{f1m81sn}. The outer overlayed circle shows
the fit outer radius, \thout, of the shell model with absorption in
the center and indicates the expected location of the forward shock
front, which is expanding into the circumstellar medium (CSM). The
inner circle shows the fit inner radius, \thin, of that shell model
and indicates the expected location of the reverse shock front. The
forward and reverse shock fronts travel in co-moving opposite
directions (see arrows) from the contact discontinuity, or contact
surface, where the ejecta hit the CSM\@.  The location of the contact
surface is shown by the dashed circle. Fingers expanding into the
shocked CSM due to the contact surface being Rayleigh-Taylor unstable
are also shown.  The smaller dashed-dotted circle indicates the
boundary between the colder, neutral ejecta and the hotter, ionized
ejecta. Broad line emission is expected from the ionized ejecta up to
the contact surface. The center of the circles is at the fit center
position (see Paper I). Typical temperatures are also indicated.
Second panel from top: The brightness profile of the composite image
above, averaged over all p.a.'s, plotted as a function of angular
radius, $\theta$.  The resolution is 0.70~mas, and the plotted
uncertainties are the standard errors of the bin values, derived from
the number of beam areas within each bin and the larger of the
standard deviation within that bin and the rms of the noise of the
background brightness.  The plotted values are correlated, especially
at small radii, because they are less than 1 beamwidth apart.  Also
indicated is the corresponding radial profile of the projected best
fit spherical shell model with uniform emissivity and an absorption
disk in the center, convolved to the resolution of 0.70~mas.  Third
panel from top: The corresponding radial profile of the projected
unconvolved shell model with absorption. Bottom panel: The radial
profile of the volume emissivity of the shell (without
absorption). The shell is limited by \thout\ and \thin.
\label{f2snmodel}}
\end{figure}

To clarify the astrophysical interpretation of the radio shell and its
relation to the optical absorption and emission by the ejecta, we
copied the left part of the VLBI image from Figure~\ref{f1m81sn} and
juxtaposed it to a sketch of a generic supernova shell model in
Figure~\ref{f2snmodel}.  The model describes a supernova expanding
into the circumstellar medium (CSM) left over from the progenitor
star. The supernova ejecta are spherically, freely expanding away from
the explosion center for a large range of radii, $r$, from that
center.  Their temperature is $\sim10^4$~K.  The ejecta in the
outermost regions hit the CSM, and a contact surface forms between the
two media at a radius, $r_{\rm cs}$. From this surface, a forward
shock, located at a radius, $r_{\rm fs}$, travels into the CSM and
heats it up to $\sim3\times10^{8}$~K. Outside the forward shock, the CSM is
unshocked and has a relatively high temperature of 10$^5$ to 10$^6$~K
(Lundqvist \& Fransson 1988; Fransson \& Bj\"ornsson 1998;
Mioduszewski, Dwarkadas, \& Ball 2001).  Also, a reverse shock travels
from the contact surface back into the ejecta to a radius, $r_{\rm
rs}$, from the explosion center and heats up the shocked ejecta to a
temperature of $\sim10^{7}$~K. The contact surface itself is
Rayleigh-Taylor unstable, and fingers of ejecta are expected to
develop and extend into the shocked CSM\@. These fingers are also
sketched in Figure~\ref{f2snmodel}.

This general scenario above was described mathematically by the
mini-shell model of self-similar expansion (Chevalier 1982a, b; see
also Fransson, Lundqvist, \& Chevalier 1996).  This model assumes
power-law density profiles for both the ejecta and the CSM, with
$\rho_{\rm ej}\propto r^{-n}$ ($n>$5), and $\rho_{\rm CSM}\propto
r^{-s}$.  With this assumption, the expansion is self-similar, and the
shock radius is given by $r_{\rm fs}\propto t^m$, where $m$ is the
deceleration parameter, which is constant and is given by
$m=(n-3)/(n-s)$.  A hydrodynamic model for SN~1993J by Mioduszewski
\etal\ (2001), based on an ejecta model by Shigeyama \etal\ (1994) (see also 
Suzuki \& Nomoto 1995),
relaxed the assumption of a power-law density profile in the ejecta.
This model described the expansion of SN 1993J in more detail, and
accounted for the deviations from self-similar expansion which were
found for this supernova (Paper II).

The Expanding Shock Front Method is based on the transverse velocity
of the Rayleigh-Taylor unstable contact surface being equal to the
largest radial velocity of the ejecta.  Dividing the latter by the
former gives a geometric determination of the distance to the
supernova and its host galaxy.  In the following we will describe how
the contact surface velocity distorted by Rayleigh-Taylor
instabilities is related to the expansion velocity of the outer
surface of the radio shell and how it compares with the maximum
velocities measured in optical lines of the ejecta.

\subsection{The origin of the radio emission}
\label{radopt}
\subsubsection{The forward and reverse shocks}

It is believed that the forward shock accelerates electrons to
ultrarelativistic velocities. These relativistic electrons then
interact with the magnetic field amplified near the Rayleigh-Taylor
unstable contact surface (e.g., Chevalier, Blondin, \& Emmering 1992;
Jun \& Norman 1996a, b). This interaction leads to radio synchrotron
radiation.  Radio emission is therefore expected to originate in the
shell region of the shocked CSM and likely also the shocked ejecta,
i.e., in a region bound by $r_{\rm rs}$ and $r_{\rm fs}$. No radio
emission is expected from the freely expanding, unshocked ejecta.
Radio emission could also originate from a central compact source
associated with the stellar remnant of the explosion, namely a neutron
star or a black hole.  Although a compact source was indeed found inside
the shell of SN 1986J (Bietenholz, Bartel, \& Rupen 2004b), no
emission from such a source has yet been detected for SN~1993J or any
other modern supernova (e.g., Bartel \& Bietenholz 2005).

How can our radio images of SN~1993J be interpreted in detail in the
context of this astrophysical model?  The radio image of SN~1993J in
Figure~\ref{f1m81sn} shows an exemplary shell of emission.  In
general, the outer angular radii, \thout\ of our geometric shell model
could be determined with an accuracy of about 1 to 2\% and, for our
latest images, the mean ratio of the outer to inner shell radii,
\thout/\thin, to within 1\% (Paper II). In Figure~\ref{f2snmodel} we
plot the fit geometric shell model as overlayed concentric circles
with radii of \thout\ and \thin. The center of the model circles is at
the explosion center within 64~\muas\ which corresponds to only about
1\% of \thout\ (Paper I). Our finding that the radio emission can be
so well fit by a spherical shell centered at the explosion center
strongly supports the expectation that the fit angular radii, \thout\
and \thin, indeed correspond to the radii of the forward and reverse
shocks, $r_{\rm fs}$ and $r_{\rm rs}$, respectively.  Other supernovae
may not show such exemplary shells of emission and would therefore not
be as good examples to show the close
relationship between the radio shell boundaries and the forward and
reverse shocks although that relationship most likely also exists
given the example of SN 1993J\@.

\subsubsection{The thickness of the radio shell and its relation to the 
forward and reverse shocks}
\label{thickness}

Both the self-similar mini-shell model and the hydrodynamic model make
predictions as to the distance between the forward and the reverse
shocks.  During the first months, when $m=0.92$ (Paper II), the ratio
between the forward and reverse shock radii is predicted to be $1.19$
for $s=2$ for the self-similar mini-shell model. For a larger
deceleration the shell thickness is predicted to be larger.  For
instance, with the mean of $m$ for the whole observing time up to
$t\sim3000$~d of $m=0.83$ (Paper II), a ratio of $1.30$\footnote{These are
average values for SN 1993J\@. But since $m$ is observed and $s$
inferred to change with time, and the evolution is in general not
self-similar (Paper II), the ratio may differ somewhat from the
predicted value.} is predicted by the self-similar mini-shell model
for $s=2$ (Chevalier \& Fransson 1994).

\begin{figure}
\plotone{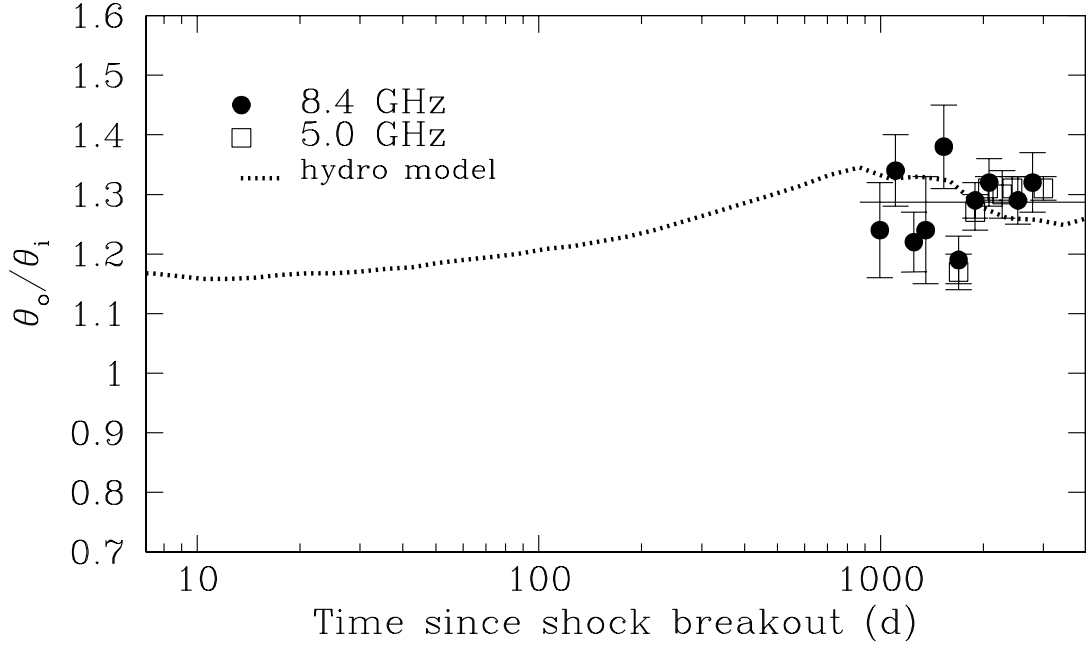} \figcaption{
The ratio \thout/\thin, which is a measure of the shell thickness, as
a function of time. The data points are our measurements at times when
the shell was sufficiently large for a thickness determination (from
Paper II).  The thin horizontal line from $t=996$~d to 2996~d gives
the mean of \thout/\thin\ = 1.29 for this time interval. The dotted
line gives the prediction from hydrodynamic simulations.
\label{f3rorilogsim}}
\end{figure}

Similar and most likely more reliable values for the shell thickness
are predicted by the hydrodynamic model.  In fact, Dwarkadas \&
Chevalier (1998) found that the shell thickness is relatively
independent of the form of the ejecta profile for a long time after
shock breakout, further supporting its robustness. In
Figure~\ref{f3rorilogsim} we plot the ratio of the forward and reverse
shock radii from a few days to several years after the explosion as
derived from the hydrodynamic simulations. The ratio increases from
1.16 between $t=10$~d and 20~d to an average of $1.29$ between
$t=1000$~d and $3000$~d. In addition we plot our measurements from
VLBI observations, \thout/\thin, determined at epochs from $t=996$~d
onward when the radio shell was large enough to allow useful
determinations.  The mean of \thout/\thin\ for epochs up to $t=2996$~d
after the explosion was $1.29\pm0.01$, with a 
slight broadening of the shell (significant only at the $1\sigma$ level,
see Paper II), in good agreement with the
models. This good agreement is additional evidence that the radio
shell is indeed bound by the forward and reverse shocks.

\subsubsection{The brightness profile of the shell and its relation 
to the forward and reverse shock} 

Below the image of the half radio shell in Figure~\ref{f2snmodel} we
plot the profile of the fit geometric shell model in several aspects.
The data points give the observed profile. The solid curve gives the
fit to the data of a (projected) shell model with uniform emissivity
between \thin\ and \thout\ and partial absorption inside \thin\ (Paper
III). The rectangular lines give the profile in the (unprojected)
cross section of the shell. Clearly, the emission and the region where
the emission originates is very well fit by our geometric model. In
Paper III we reported on the steepness of the outer edge of the
profile and found that it is indeed consistent with being an ideally
sharp edge as in our model. However softer boundaries are not
excluded. Nevertheless, the measured brightness profile can be taken
as further evidence that the radio shell is indeed bound by the
forward and reverse shocks, and that the parameters \thin\ and \thout\
therefore most likely correspond closely to $r_{\rm rs}$ and $r_{\rm
fs}$.

On the right side of Figure~\ref{f2snmodel} we sketch the minishell
model and complete the concentric circles by indeed assuming here and
hereafter that \thin\ and \thout\ correspond exactly to $r_{\rm rs}$
and $r_{\rm fs}$.

\subsection{The origin of the optical emission} 
\label{opt}
While optical continuum emission of the supernova is expected to
mostly originate, at least at early times, from the photosphere in the
central region of the expanding gas, optical emission and absorption
in broad spectral lines is expected to arise in the gas of the hot
ejecta, above the cooler and mostly neutral ejecta (Figure~\ref{f2snmodel}. 
In particular,
emission was discussed as coming from a) the freely expanding ejecta
and being caused by the heating and ionizing radiation from the
interaction region, as well as from b) the shocked ejecta and being
linked to radiative cooling (Chevalier \& Fransson, 1994, and
references therein). The ionizing luminosity and the density of the
gas both change with time and give rise to evolving spectra.  Blending
of lines is common, and since the relative luminosities change with
time, the degree of blending can change as well. The highest expansion
velocity in absorption and emission lines is expected from the shocked
gas of the envelope of the progenitor star, in a region that extends
outwards to the Rayleigh-Taylor unstable contact surface at radius
$r_{\rm cs}$.  The highest {\em observed} optical expansion velocity
could be smaller.  No optical line emission is expected from the
region of shocked CSM, i.e., beyond the contact surface between
$r_{\rm cs}$ and $r_{\rm fs}$, since the temperature there is expected
to be $\sim3\times10^{9}$~K.

\subsection{The contact surface}
\subsubsection{The location of the contact surface within the radio shell}
The location of the contact surface in relation to the forward and
reverse shocks, i.e., to the outer and inner radio shell surfaces, is
estimated by combining VLBI measurements of the radio shell thickness
with circumstellar interaction models.  Both the self-similar
mini-shell model and the hydrodynamic model make predictions as to the
distance between the contact surface and the reverse shock.  During
the first months, when $m=0.92$, the ratio between the contact surface
and reverse shock radii is predicted to be $1.0065$ for $s=2$ by the
self-similar mini-shell model. For a larger deceleration, e.g., for a
mean of $m=0.83, s=2$ as taken before for a comparison
(\S~\ref{thickness}), the ratio is $1.03$ (Chevalier \& Fransson
1994). The hydrodynamic model predicts a ratio similarly close to
unity during the first month which then increases to a mean value of
about 1.04 between $t\sim1000$ and $\sim3000$~d.

We parametrized \thin\ for the range of our radio and the optical
observations by \thin= $A_{\rm rf} t^{m_{\rm rf}}$ \thout, where $t$
is the time since explosion in days. With $A_{\rm rf}= 0.914$ and
$m_{\rm rf}=-0.0217$, this parametrization gives a time-varying shell
thickness consistent with the hydrodynamic simulations with, e.g.,
\thout/\thin\ = 1.16 at $t=15$~d and 1.29 at $t=3000$~d (see
Figure~\ref{f3rorilogsim}). These values are also fairly consistent
with the predictions from the self-similar mini-shell model. (We also
parametrized the angular radius of the contact surface by \thc =
$A_{\rm cf} t^{m_{\rm cf}}$ \thout\ with $A_{\rm cf}= 0.904$ and
$m_{\rm cf}=-0.0150$ to give \thc\ being 0.65\%\ and 4\%\ larger than
\thin\ at $t=15$~d and 2500~d, respectively, as predicted by the above
models.
 
\subsubsection{Distortion of the contact surface by Rayleigh-Taylor instabilities}

However, since the contact surface is decelerating, it is expected to
be Rayleigh-Taylor unstable (Gull 1973; Chevalier 1982b), and that
with time fingers of shocked ejecta will extend into the shocked
CSM\@.  A linear analysis of self-similar solutions as well as
two-dimensional numeric hydrodynamic computations have shown that
Rayleigh-Taylor instabilities are limited by Kelvin-Helmholtz
instabilities but can still grow in length to generally 30 to 50\% of
the width of the shocked shell (Chevalier \etal\ 1992; Dwarkadas
2000).  Under particular circumstances, when vortices are amplified by
the magnetic field, it was shown that the fingers may even reach
slightly beyond the outer shock, which is otherwise assumed to be
spherically symmetrical (Jun \& Norman 1996a, b). Also, for fast
shocks with efficient particle acceleration and resulting high
compression ratios, the width of the shocked shell was found to shrink
considerably so that fingers, without growing much larger, could also
reach and even slightly exceed the average forward shock radius
(Blondin \& Ellison 2001).

\subsubsection{The fastest parts of the ejecta}
The ejecta with the highest velocity are therefore located in the
evolving Rayleigh-Taylor fingers stretching out beyond $r_{\rm cs}$
toward the forward shock at $r_{\rm fs}$.  The fingers are expected to
be less developed a few days after shock breakout and more developed a
few years thereafter. Thus, the highest possible velocities in optical
lines are those from the front of the Rayleigh-Taylor
fingers\footnote{Note, that whether such line emission is observable
depends on the physical condition in the fingers.} which, depending on
evolution time, stretch out to between $r_{\rm cs}$ and $r_{\rm fs}$
with an assumed mean, $r_{\rm x}=\onehalf(r_{\rm fs}+r_{\rm cs})$,
which corresponds to an angular extent of \thx.

\section{THE EXPANSION}
\label{exp}

\subsection{The angular expansion velocities of different surfaces of and within the radio shell}
\label{radvel}

\begin{figure}
\plotone{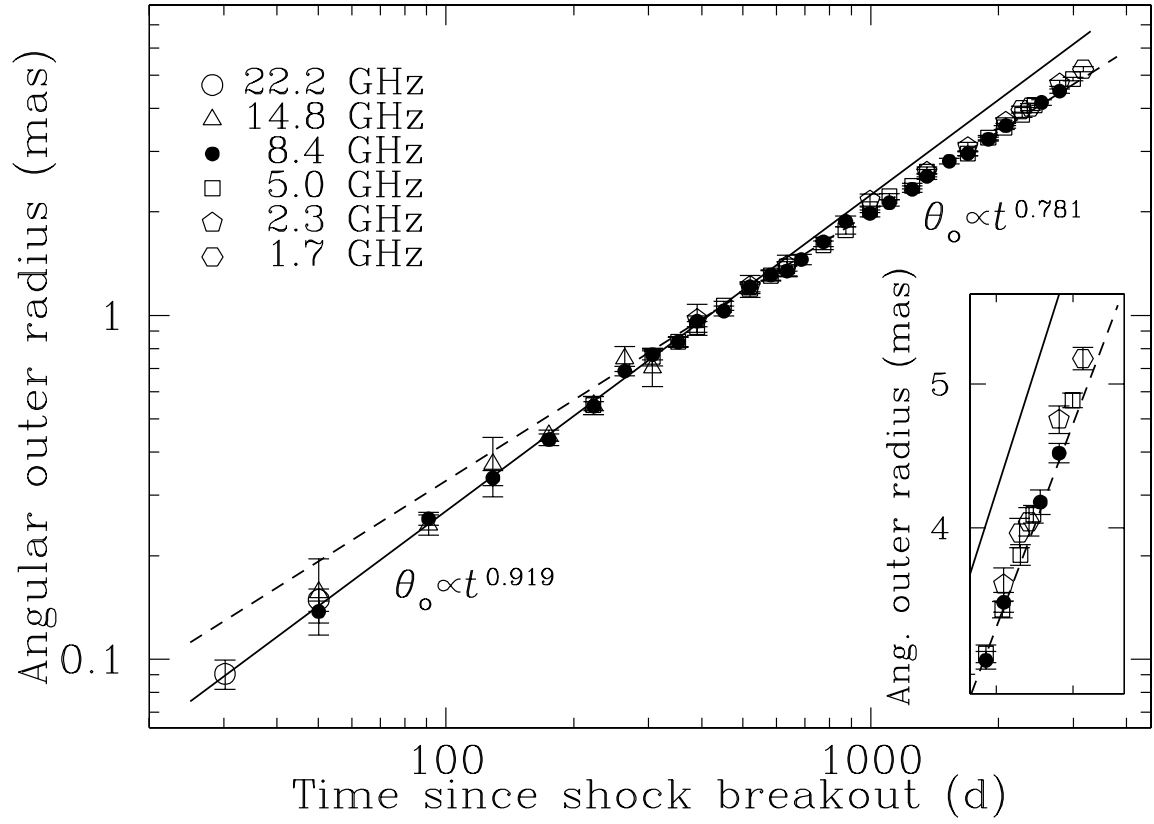}
\figcaption{The angular outer radius, \thout, of the SN
1993J shell model as a function of time since shock breakout. Taken from Paper II.
\label{f4exp}}
\end{figure}

We used the values of $\thout$ obtained from our VLBI observations to
compute the angular expansion velocity of the outer surface of the
radio shell, i.e., of the forward shock, $\dotthout$, as a function of
time.  The values of \thout\ were already listed and plotted in
Paper~II, but for the convenience of the reader we repeat the plot of
\thout\ as a function of time here in Figure~\ref{f4exp}.  It can be
clearly seen that the outer surface of the radio shell is only
slightly decelerated up to $t\sim300$~d and from then on more strongly
decelerated. Further, as can be seen from the inset of
Figure~\ref{f4exp}, the deceleration changes significantly even after 
$t\sim2000$~d (see also Paper II).

Because the expansion has an approximate power-law form, a linear fit
to the values of $\thout$ over a time interval would produce a
biased estimate of $\dotthout$ at the midpoint of the interval.  To
avoid this bias, we computed $\dotthout(t)$ by first fitting a running
solution of the form $\thout = A t^m$ to the data points near $t$,
where we fit both the deceleration parameter, $m$, and the angular
size scale, $A$.  Our running solution is equivalent to a running
mean, but instead of computing the weighted average in a shifting
interval we computed the weighted linear fit in a shifting interval in
a log-log diagram (see Paper~II for more details).  As our value of
$\dotthout(t)$, we take the derivative of this fitted function, $A m
t^{m - 1}$, evaluated at the midpoint of the fitting interval on a
logarithmic scale\footnote{In particular, we computed a ``running
solution,'' $\theta = A_N t^{m_N}$, for each $N$, from the values of
$\thout$ for 8 consecutive observing epochs from $t_N$ to $t_{N + 7}$.
We started with the data from the first epoch at $t_1$ to the eighth
epoch at $t_8$ for the first solution ($N = 1$), and continued till
$t_{N+7}$ was the last epoch.  For each $N$, we fit for the values of
$A_N$ and $m_N$.  The time at which these fitted values of $A_N$ and
$m_N$ are taken to apply, $\bar{t_N}$, is the geometric mean of the
start and end times of the segment, i.e., $\log{\bar{t_N}} =
{1\over{2}} (\log{t_{N}}+\log{t_{N+7})}$.  At $\bar{t_N}$ we take the
angular velocity to be given by $\dotthout(\bar{t_N}) = A_N m_N
\bar{t_N}^{m_N - 1}$.}.

\begin{figure}
\plotone{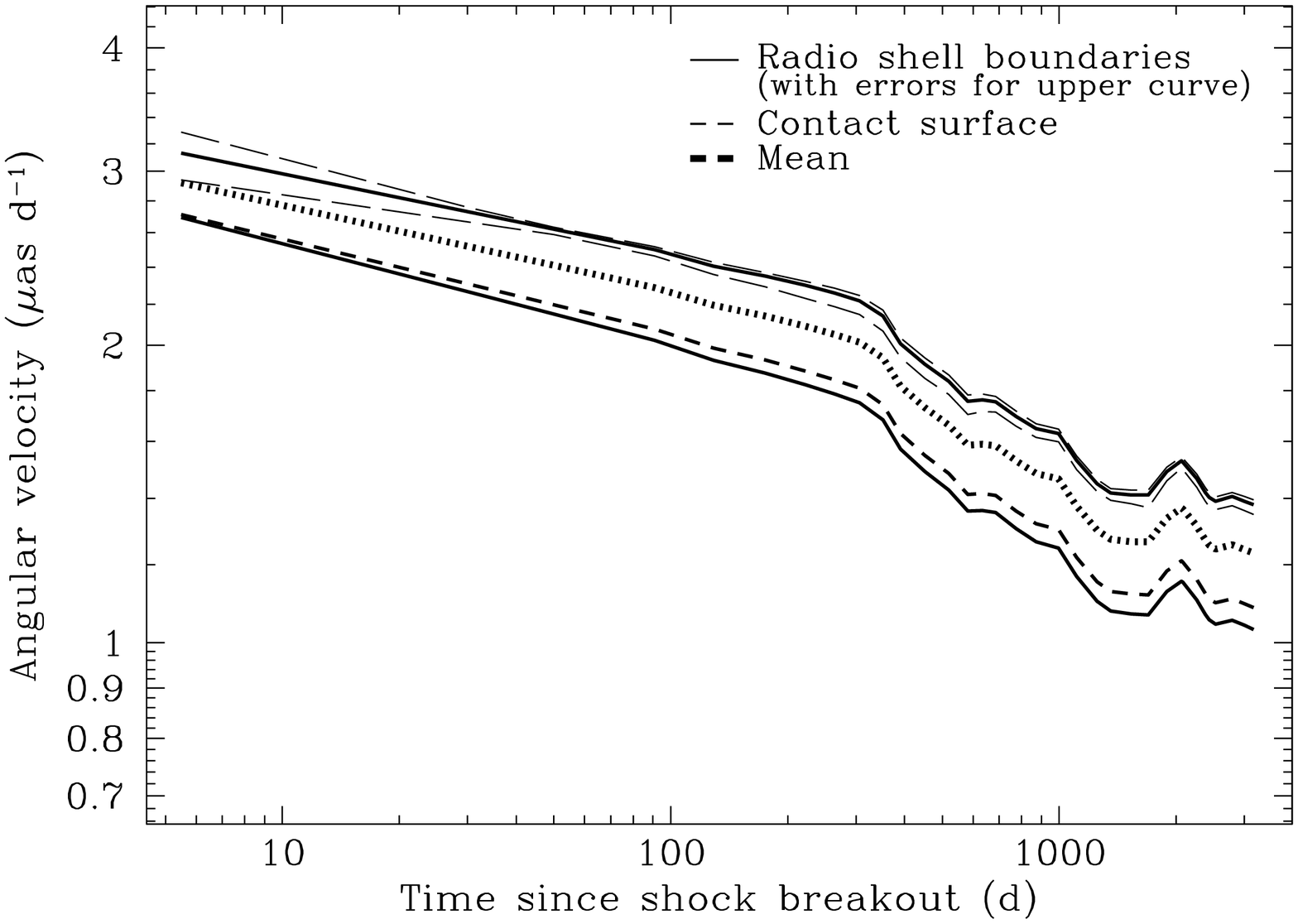}
\figcaption{The time derivatives of the angular radii of the outer 
(\dotthout, upper solid curve) and inner (\dotthin, lower solid curve) surfaces of the
radio shell, the contact
surface (\dotthc, curve with short dashes), and the mean between the
outer surface and the contact surface (\dotthx, curve with dots). The
thin curves with long dashes on both sides of the upper solid curve
indicate the (asymmetric) standard error of 1 to 5\% of \dotthout\ as
a function of time (see text).
\label{f5radvellogm}}
\end{figure}

We extrapolated the angular velocity values slightly beyond the mean
times of the first and last running solutions to cover the time range
from $t\sim5$ to $\sim3000$~d.  We plot the values of \dotthout\ and
connect them by lines in Figure~\ref{f5radvellogm}. We also estimate
standard errors from the power-law fits. These errors are between 1
and 5\% of \dotthout.  We plot them also in Figure~\ref{f5radvellogm}.
We further compute the angular velocity of the inner surface of the
radio shell, i.e. of the reverse shock, \dotthin, the velocity
of the contact surface, \dotthc, and the mean between the velocity of the contact
surface and that of the forward shock, \dotthx, and also plot them in
Figure~\ref{f5radvellogm}.  The ratios of the parametrized velocities,
\dotthout/\dotthin\ and \dotthout/\dotthc, are comparable to those from the
hydrodynamic simulations after the latter are smoothed over
appropriate time intervals.\footnote{The ratio, $r_{\rm fs}/r_{\rm
rs}$, predicted by the hydrodynamic simulations
(Figure~\ref{f3rorilogsim}) shows an oscillation at late times which
corresponds to an oscillation of the ratio $\dot r_{\rm fs}/\dot
r_{\rm rs}$.  However, our data are also consistent with our simple
parametrized function which does not oscillate. Our ratios of the
parametrized velocities at late times are similar to the mean of the
high and low values of the oscillation of the velocity ratios from the
hydrodynamic simulation.}

\subsection{The radial expansion velocities of the optical line-emitting gas} 
\label{optvel}
\subsubsection{The data}
We list the maximum expansion velocities found in optical lines of
hydrogen, oxygen, and sodium in Table~\ref{t1vopt}.  We obtained the
earliest velocities from measurements by Finn \etal\ (1995) who
reported 17,800 declining to 16,600 \kms\ from $t=7$ to 10~d after
shock breakout for the {\em minimum} of the H$\alpha$ $\lambda$6563
absorption trough. The maximum velocities are larger. Since an
absorption profile was not published, we needed to calculate the
maximum velocities on the basis of absorption profiles at overlapping
times published by others. For $t=10$ to 19~d, maximum velocities of
$18,000$ \kms\ were obtained from the blue edge of the H$\alpha$
absorption trough (Trammell \etal\ 1993; Lewis \etal\ 1994).  Since at
$t=10$~d the maximum velocity is 1400 \kms\ higher than the velocity
at the absorption minimum, we added 1400 \kms\ to Finn \etal's values
to obtain the maximum velocities also for $t=7$ to 9~d.  The highest
velocity is 19,200 \kms\ at $t=7$~d.

At later times the absorption trough first shrank, most likely due to
blending at the blue side with the rising [OI] $\lambda\lambda$ 6300,
6364 emission doublet\footnote{Patat, Chugai, \& Mazzali (1995) report
a sharp decrease of the radial velocity of the absorption minimum
in the H$\alpha$ profile from $\sim13,500$~\kms\ to $\sim9500$~\kms\
between $t=15$~d to $\sim50$~d. From the spectra in this time range
(Matheson \etal\ 2000a) it appears that the sharp decrease is caused
by the [OI] $\lambda\lambda$ 6300, 6364 emission doublet filling in
the blue side of the H$\alpha$ trough even at these early
times.}. Then the trough disappeared due to H$\alpha$ becoming
optically thin (see, e.g., Houck \& Fransson 1996), and the maximum
velocity for H$\alpha$ could only be derived from the H$\alpha$
emission lines.

Several authors have measured emission lines in SN~1993J, e.g., Lewis
\etal\ 1994; Spyromilio 1994; Filippenko, Matheson, \& Barth 1994;
Clocchiatti \etal\ 1995; Patat \etal\ 1995; Houck and Fransson 1996;
Matheson \etal\ 2000a, b; Fransson \etal\ 2005. Of these, Patat \etal\
(1995) and Matheson \etal\ (2000a, b) reported measurements over the
longest time intervals.  Patat \etal\ (1995) plot representative
H$\alpha$ profiles and list RVZI (red velocity at zero
intensity) values for all epochs of their
observations. We list these values in Table~\ref{t1vopt}.  Matheson
\etal\ (2000a) plot the profiles for all epochs of their observations
and list (Matheson \etal\ 2000b) BVZI (blue velocity at zero
intensity) as well as RVZI values. We considered the possibility of
asymmetries of the line profiles that could indicate absorption
effects or possible biases in the determination of the maximum
velocities and therefore determined from the profiles consistently the
BVZI and RVZI values\footnote{We calculate the velocity at zero intensity by
extrapolating the side of the line through the noise to the zero
intensity baseline of the spectrum.} as well as the BVHI (blue velocity at half
intensity) and RVHI (red velocity at half intensity)
values\footnote{Matheson \etal\ (2000b) list H$\alpha$ BVZI and RVZI
values but not BVHI and RVHI values.  For consistency we did not use
Matheson \etal's BVZI and RVZI values but rather determined these
values and the corresponding BVHI and RVHI values from the line
profiles given by Matheson \etal\ (2000a). Matheson \etal's (2000b)
values do not differ from our determined values significantly, except
at $t=523$ and 553~d where we think that their BVZI values are
underestimates. These underestimates are likely caused by a
misinterpretation of the baseline level, which we believe is taken too
high due to the still present blend of the O[I] $\lambda\lambda$ 6300,
6364 emission doublet (see \S~\ref{comparison}).}.  These values are
also listed in Table~\ref{t1vopt}.

\subsubsection{The maximum H$\alpha$ velocities}
\label{allhalpha}
\begin{figure}
\plotone{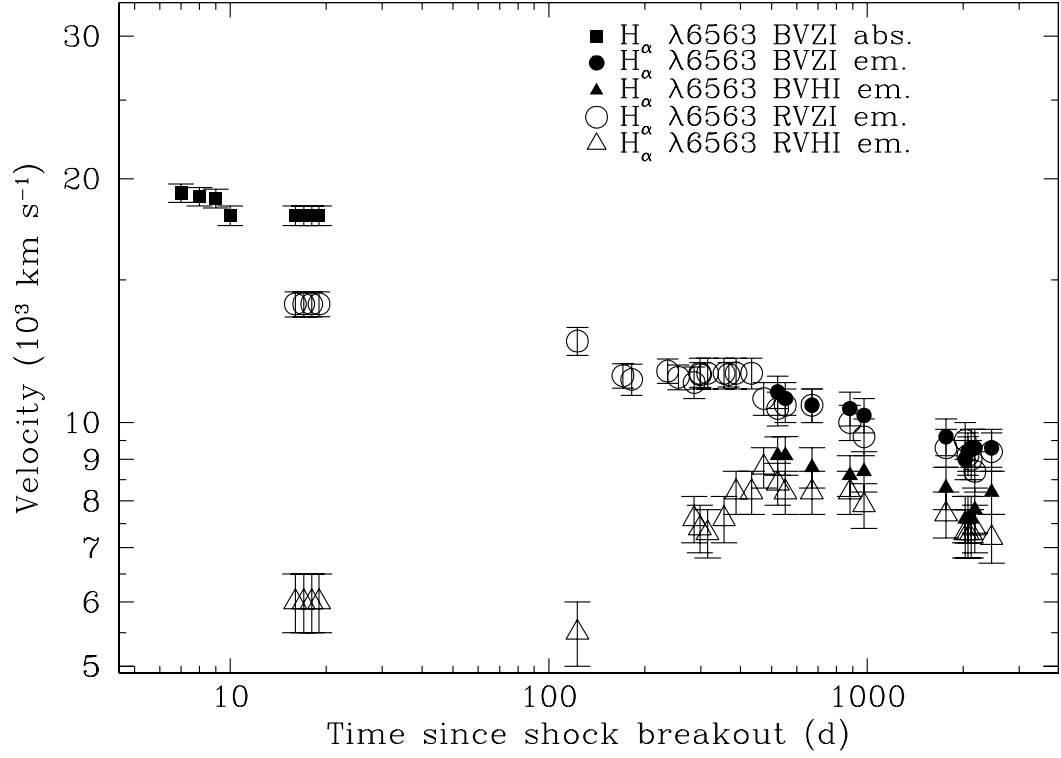}
\figcaption{The observed expansion velocities from the
width of the H$\alpha$ line from Table~\ref{t1vopt} as a function of
time.  The symbols represent the BVZI (blue velocity at zero
intensity) values from the absorption trough of the line (filled
square), and the BVZI (filled circle), BVHI (blue velocity at half
intensity, filled triangle), RVZI (red velocity at zero intensity,
open circle) and RVHI (red velocity at half intensity, open triangle)
values, all from the emission line.
\label{f6halphaalllogmd}}
\end{figure}

For a supernova like SN 1993J, hydrogen gas is a dominant constituent
of the outer regions of the freely expanding ejecta and the shocked
ejecta. We therefore plot the BVZI, BVHI, RVZI, and RVHI values of the
H$\alpha$ profile from Table~\ref{t1vopt} as a function of time in
Figure~\ref{f6halphaalllogmd}.

It is apparent that the maximum H$\alpha$ velocities decrease with
time. In particular, the BVZI decreases at a rate similar to that of
\dotthx\ with the BVZI value at $t=2454$~d being about half as large
as that at $t=7$~d.  The BVZI values at early times, between $t=7$ and
$19$~d, are determined from the blue edge of the absorption
trough. They are between 19,200 and 18,000 \kms, the largest
velocities measured for SN~1993J at any time. After that time, the
O[I] $\lambda6300$, $6364$ line appears and steadily grows in
intensity, rendering impossible a measurement of the H$\alpha$ BVZI in either
absorption or emission or of the BVHI in emission.  Only at
$t\geq523$~d could the velocities on the blue side of the H$\alpha$
line be measured again.  Figure~\ref{f6halphaalllogmd} further shows
that the earliest RVZI values, from $t\sim16$ to $\sim19$~d, are
$\sim$20\% smaller than the corresponding BVZI values. At later times,
however, the RVZI values agree within the errors with the BVZI
values. Furthermore, a relatively large discrepancy exists between the
RVZI and RVHI values at early times, the latter being almost 60\%
smaller than the former.  This discrepancy decreases at later times
due to a steepening of the red side of the H$\alpha$ line profile.  At
$t\gtrsim500$~d, the RVZI and RVHI values are different by only
$\sim$20\% and match the corresponding values on the blue side,
indicating a largely symmetric profile with steep red and blue sides.

These characteristics of the H$\alpha$ profile at times up to
$t\sim500$~d indicate that the red side may be biased by 
absorption, likely by dust mixed in with the ejecta. The occurrence of
dust in supernovae and absorption by it was discussed by, e.g.,
Fransson \etal\ (2005), Gerardy \etal\ (2000), and Deneault, Clayton,
\& Heger (2003).  The maximum velocity of the H$\alpha$ line emitting
gas is therefore best revealed by the BVZI (absorption and emission)
values at all times and the RVZI (emission) values for
$t\gtrsim500$~d\footnote{Patat \etal\ (1995) discussed a possible
blending of the red side of the H$\alpha$ emission profile at early
times ($t$=255~d) with an unidentified feature reported to be present
in Type Ib/c supernovae but limited the emission to not more than 30\%
of that of H$\alpha$. However, this feature, visible in Patat \etal's
(1995) example, cannot be seen in the spectrum of SN 1993J\@.
Further, Chevalier \& Fransson (1994) computed that another line, [N
II] $\lambda$6548-6583, could be blended with the H$\alpha$ line.
However, at $t=2$~yr its luminosity would be only $\sim$5\% of that of
the H$\alpha$ line and at $t=5$~yr only 20 to 25\%. These limits and
the time frame of the occurrences of these possible blends together
with the near consistency of the RVZI with the BVZI values for
$t\gtrsim500$~d are strong indications that our values are not
significantly affected by blending with such lines.}.

\subsection{Radio versus optical deceleration}
\label{radoptdecel}
How well does the deceleration of the radio shell match that derived
from the maximum velocities from the optical lines?  In
Table~\ref{t2mfit} we compare the deceleration parameter of the radio
shell, $m$, with that of the line absorbing or emitting gas, $m_{\rm
opt}$, for the lines discussed above and for different time ranges. In
particular we compute $m_{\rm o}$ from \thout$\propto t^{m_{\rm o}}$,
$m_{\rm c}$ from \thc$\propto t^{m_{\rm c}}$, $m_{\rm x}$ from
\thx$\propto t^{m_{\rm x}}$, and $m_{\rm opt}$ from $v_{\rm
opt}\propto m_{\rm opt} t^{(m_{\rm opt}-1)}$, where $v_{\rm opt}$ is
the maximum velocity of the line absorbing or emitting gas.

We first computed the deceleration of the radio shell and that of the
H$\alpha$ line absorbing and emitting gas for the total time
range. Here we had to consider the uneven sampling of the radio and
optical data. Since the deceleration is changing with time, a fit with
a single deceleration parameter would depend strongly on the weighting
scheme. In fact, on a logarithmic scale, the radio data for instance
were only sparsely sampled at early times and more densely sampled at
later times. If the non-uniform sampling is ignored, then as reported
in Paper II, $m_{\rm o}=0.827\pm0.004$ for a fit weighted only with
the data uncertainties as given in Table II of Paper II. However, such
a fit represents a strong mismatch to the earliest data. For a better
comparison with the optical data including the earliest data from
$t=7$ to 19 d we increased the weighting of the early radio data by
forcing the fit through the earliest radio data point at $t=30$~d. For
the optical data such weighting was not necessary mostly due to the
large gap at intermediate times.

The resulting ``average'' deceleration parameter for the H$\alpha$ gas 
is equal within the errors combined in quadrature to that
of the outer surface of the radio shell and that of the surface between the
latter and the contact surface (solution 1 in Table~\ref{t2mfit}). The
equality within the errors also holds for the early time interval
(solution 2 in Table~\ref{t2mfit}).  For the late time interval, the
H$\alpha$ line absorbing and/or emitting gas is slightly, but
significantly, less decelerated than the radio shell with a difference
in the deceleration parameters of 0.092 (4$\sigma$) and 0.099
(4$\sigma$) between the optical line emitting gas on the one hand and
the outer surface of the radio shell and the surface between the later
and the contact surface, respectively, on the other hand (solution 3
in Table~\ref{t2mfit}).

Somewhat larger discrepancies in the sense of less deceleration are found for
the O[III] and Na I line emitting gas.  The largest discrepancy is
found for the H$\beta$ line-emitting gas, where no significant
deceleration was measured. This is due to the relatively large spike
of the H$\beta$ velocities at $t\sim$2000~d (see Table~\ref{t1vopt}).

\section{THE DETERMINATION OF THE GEOMETRIC DISTANCE TO SN~1993J AND M81} 
\label{dist}

\subsection{The distance solution with statistical errors only.}

As we have argued, the most reliable measurements of the maximum
velocity from optical lines spanning the longest time are those from
the blue edge of the H$\alpha$ absorption trough at early times of $t
= 7$ to $19$~d and of the H$\alpha$ emission profile later on. We
therefore take these H$\alpha$ BVZI values (Table~\ref{t1vopt}) and
fit to them \dotthx, which is the mean between \dotthc\ and \dotthout,
taken to account for the Rayleigh-Taylor fingers. More precisely, we
interpolated \dotthout$(t)$ to the times, $t_j$, of the optical
data, $v_{\rm opt}(t_j)$, and computed the corresponding
\dotthx$(t_j)$ values to get distance estimates,
$$D_j = {{v_{\rm opt}(t_j)} \over {\dotthx (t_j)}}$$
for each $t_j$. We then took the weighted mean of the $D_j$ values,
and its uncertainty, as the solution (solution 1 in
Table~\ref{t3dfit}) for the distance, $D$, to SN~1993J and therefore
to M81\footnote{ If we ignore that the deceleration parameters for the
radio shell and the hydrogen gas are changing with time, and take
instead the ``average'' deceleration parameters, $m_{\rm x}$ and
$m_{\rm opt}$, from solution 1 in Table~\ref{t2mfit}, the distance is
given as $D={{v_{\rm opt}} \over {\dotthx}}$, that is $D={A_{\rm
opt}\over{A_{\rm x}}} {{m_{\rm opt}}\over{m_{\rm x}}} t^{m_{\rm
opt}-m_{\rm x}}$ with $v_{\rm opt}=A_{\rm opt}m_{\rm opt}t^{m_{\rm
opt}-1}$ and \thx$=A_{\rm x}t^{m_{\rm x}}$.  We get slightly lower
values for D between 3.83 Mpc for $t=10$~d and 3.91 Mpc for
$t=2000$~d. This bias is caused by the inferior fit to the radio data
and the resulting bias of slightly higher predicted velocities at
early and late times where the optical data were obtained. We do not
consider these solutions further.}.

$$D=3.96\pm0.05~{\rm Mpc}.$$

\begin{figure}
\plotone{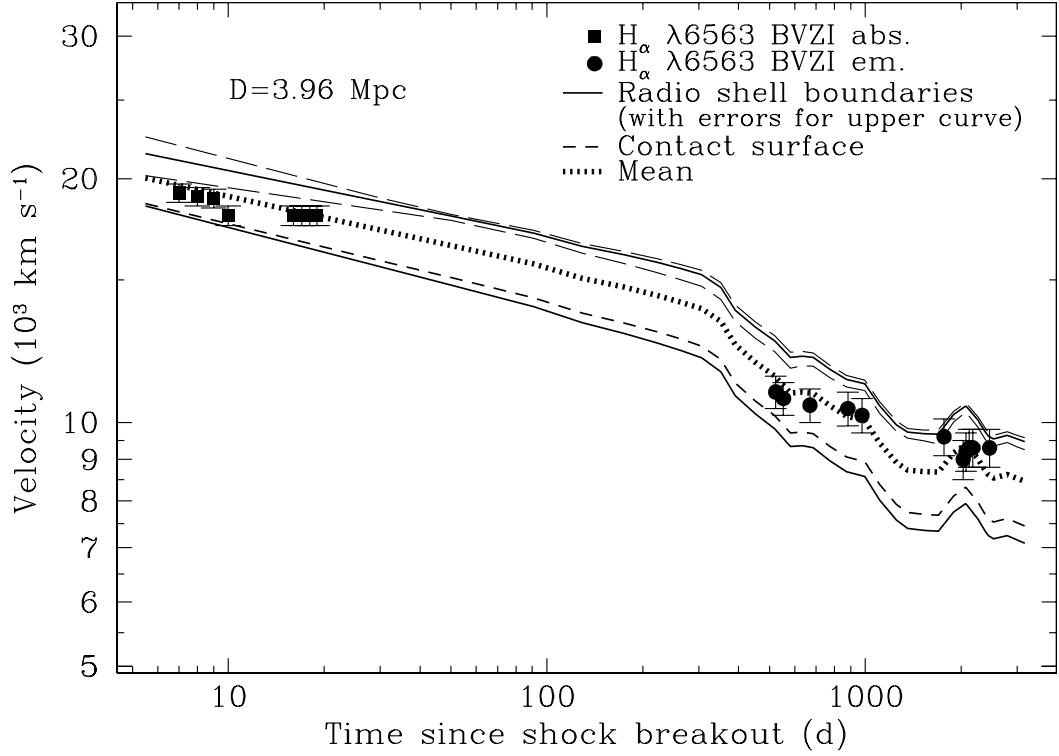}
\figcaption{As in Figure~\ref{f6halphaalllogmd} but now showing only the 
BVZI H$\alpha$ values to which  $D$\dotthx\ (Mean) was fit to determine the 
distance, $D$. In addition the velocity curves, $D$\dotthout, $D$\dotthin, 
$D$\dotthc, and $D$\dotthx\  are plotted (see Figure~\ref{f5radvellogm}) 
for the best fit distance of $D=3.96$~Mpc.
\label{f7halphablueradvellogmd}}
\end{figure}

We plot the
H$\alpha$ emission and absorption BVZI values and the radio shell
velocity curves including the velocity curve for the contact surface
and the mean between it and the outer surface (see
Figure~\ref{f5radvellogm}) for the newly determined distance of
3.96~Mpc in Figure~\ref{f7halphablueradvellogmd}.

\subsection{Sensitivity study}
To investigate how much our distance determination depends on our
choice of the optical velocity measurements, we solved for
$D$ by using subsets of the H$\alpha$ velocities, velocities from
other lines, and combinations of velocity sets, all from
Table~\ref{t1vopt}. We list the solutions also in Table~\ref{t3dfit}.

\subsubsection{Distance with H$\alpha$ velocities}
In particular, we used only the velocities from the blue edge of the
H$\alpha$ absorption profile, BVZI abs. (solution 2 in
Table~\ref{t2mfit}), from the blue edge of the H$\alpha$ emission
profile, BVZI em. (solution 3), and from the red edge of the H$\alpha$
emission profile, RVZI, from $t=523$ to 2454~d (solution 4).  These
velocity values were already plotted in Figures~\ref{f6halphaalllogmd}
and \ref{f7halphablueradvellogmd}. These distance estimates straddle
the one in solution 1, differing from it by not more than 0.05 Mpc or
1.3\% (0.6$\sigma$, combined statistical uncertainty\footnote{Strictly
speaking, combining the uncertainty in quadrature as we did gives only
approximately the statistical uncertainty of the difference, since the
data are overlapping and therefore not statistically independent.}),
having somewhat larger errors, and $\chi^{2}_\nu$ values closer to
unity apart from $\chi^{2}_\nu =0.20$ in solution 2.

\subsubsection{Distance with H$\beta$ velocities}
\begin{figure}
\plotone{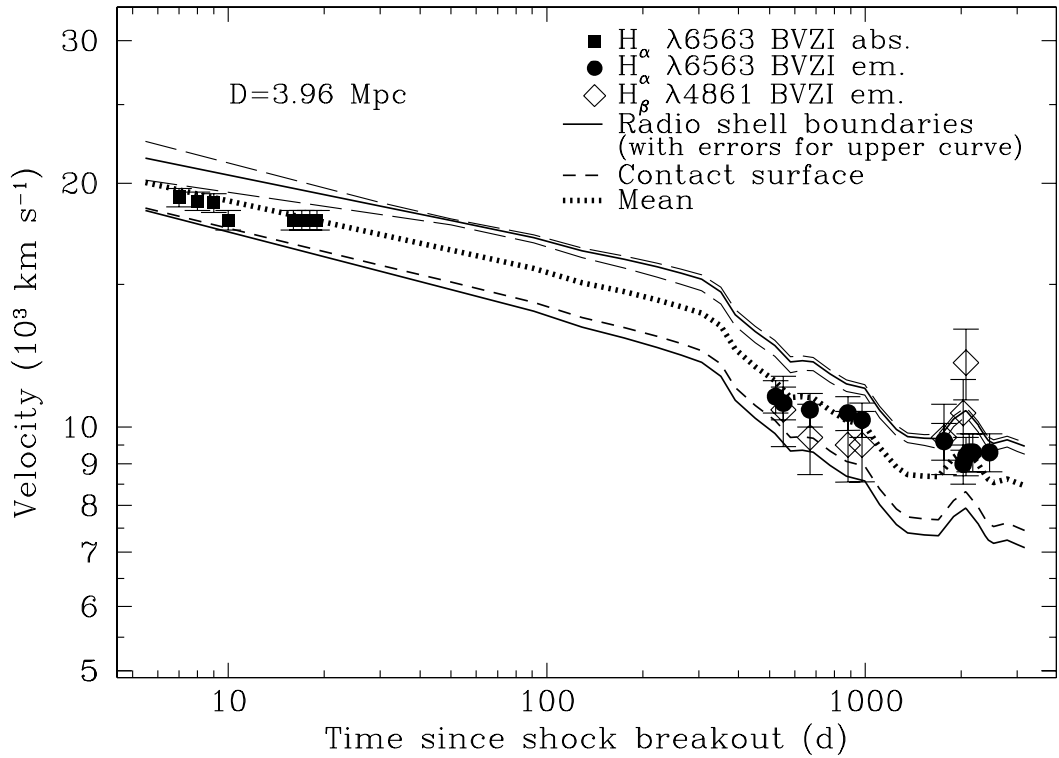}
\figcaption{As in Figure~\ref{f7halphablueradvellogmd} but now also 
with the BVZI values of the H$\beta$ emission line from Table~\ref{t1vopt}. 
\label{f8hbetaradvellogmd}}
\end{figure}

We further used the velocities from the blue edge of the H$\beta$
$\lambda$4861 emission profile (BVZI) from $t=553$ to 2454~d (solution
5).  We plot the H$\beta$ BVZI values and for comparison, the
H$\alpha$ BVZI values with the radio velocity curves for the distance
of 3.96~Mpc (solution 1) in Figure~\ref{f8hbetaradvellogmd}.  The
H$\beta$ BVZI values from $t\sim500$ to $\sim1000$~d are $\sim$10\%
smaller than the corresponding H$\alpha$ BVZI values, however they get
larger with time and at $t\sim2000$~d, even exceed the H$\alpha$ BVZI
values, just at the time when the outer radio shell velocity is
spiking up.  It is interesting to note that this behavior is reflected
in a bump in the lightcurve at X-rays (Zimmermann \& Aschenbach
2003). The distance with these values only (solution 5) is within 0.06
Mpc or 1.5\% (0.3$\sigma$, combined statistical uncertainty) equal to
the distance from solution 1.

\subsubsection{Distance with all hydrogen velocities combined}
Since the hydrogen gas constitutes a large fraction of the outer parts
of the ejecta, we combined all the BVZI and RVZI values from the
H$\alpha$ and H$\beta$ profiles used for solutions 1, 4, and 5 and
again solved for the distance.  The resulting distance (solution 6) is
only 0.01 Mpc or 0.3\% (0.2$\sigma$, statistical uncertainty from
solution 1) smaller than the distance from solution 1.

\subsubsection{Distance with O[III] and Na I velocities}
\begin{figure}
\plotone{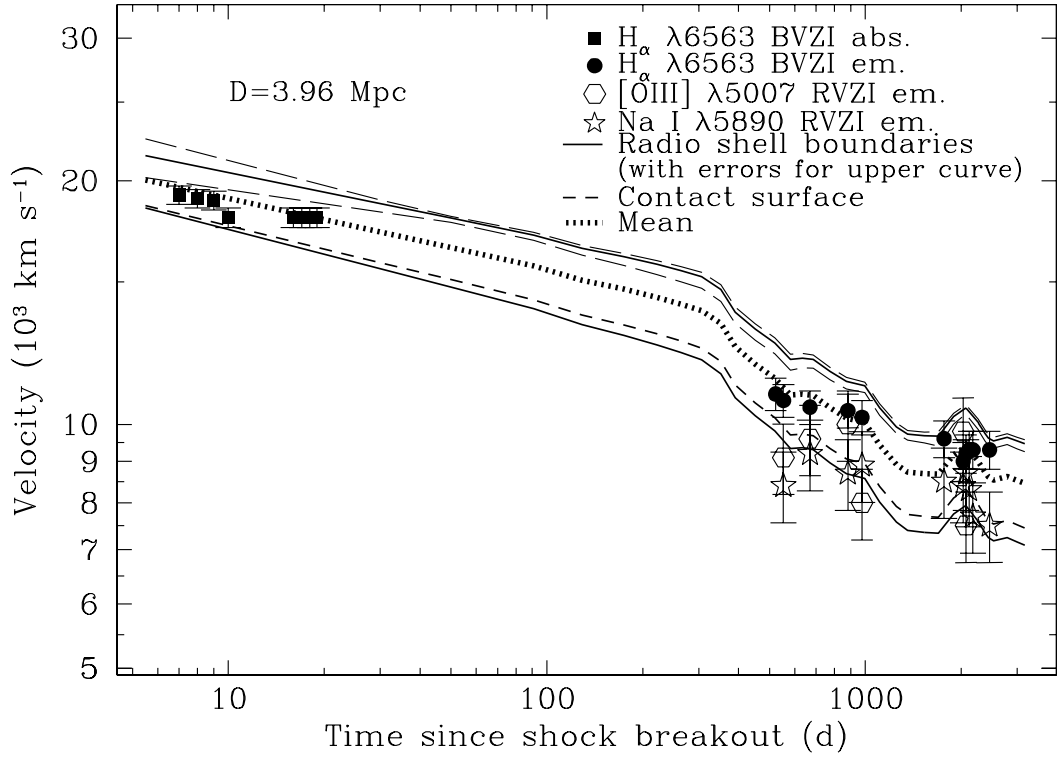}
\figcaption{As in Figure~\ref{f7halphablueradvellogmd}, but now also with 
the RVZI values of the O[III] and Na I  emission lines from Table~\ref{t1vopt}. 
\label{f9otherradvellogmd}}
\end{figure}

In Figure~\ref{f9otherradvellogmd} we plot the O[III] $\lambda$5007
and Na I $\lambda$5890 RVZI values from Table~\ref{t1vopt} and again,
for comparison, the H$\alpha$ BVZI values and the radio velocity
curves for the distance of 3.96~Mpc (solution 1).  The scatter in the
OIII RVZI values is relatively large, extending over more than
2000 ~\kms\ or $\sim$20\%.  The scatter of the Na I RVZI values is
smaller.  Almost all of the velocity values from these two spectral
lines are clearly smaller than the corresponding values from the
H$\alpha$ line, on average by 12\%. Correspondingly, the distance
estimates (solutions 7, 8) are also 12\% smaller than the estimate of
solution 1.  However, we think that these estimates do not reflect the
real distance of M81 but rather the smaller maximum radii and
velocities of the observed O[III] and Na I line gas in the ejecta.

\subsection{Systematic errors\label{systematic}}
The systematic errors of our distance estimate mostly depend on how
reliably the radio and optical velocities can be equated. In this
context, the degree of isotropy of the expansion and the spatial
relation between the radio and optical emission regions are important
factors.  We have identified seven items concerning systematic
uncertainties and elaborate on them below.

\begin{trivlist}
\item{1.} {\em Large scale anisotropy of the radio expansion:}

The degree of isotropy of the expansion of the outer radio shell, at
least in the plane of the sky, can be determined most directly by
measuring the angular expansion of individual segments of the
supernova from a fixed reference point in the frame of the host
galaxy. Such measurement was made for SN~1993J relative to M81$^*$,
the core of the nuclear region of the galaxy, and resulted in a limit
on anisotropic motion in any direction in the plane of the sky of
5.5\%\ (Paper I). More stringent limits can be set on the ellipticity
of the supernova's projection. For our composite image in
Figure~\ref{f1m81sn}, an ellipse fitted to the 20\% contour is
circular even to within 1\%.  We take the mean of these two anisotropy
measurements of 3\% for the 1$\sigma$ contribution to the systematic
error.

\item{2.} {\em Small scale anisotropy of the radio expansion:}

As we reported in Paper III, the composite image in
Figure~\ref{f1m81sn} shows an apparent small protrusion to the
southwest.  Such protrusions may result from Rayleigh-Taylor
instabilities of the contact surface, which causes vortices through
the region of the shocked CSM between \thc\ and \thout.  Such vortices
are expected to show enhanced polarization. No significant
polarization has however been found in several of the images we
analyzed, with a 3$\sigma$ limit of 9\% of the image peak averaged
over the beam size. Further, the significance of the protrusion itself
is marginal. The noise-corrected 1-$\sigma$ upper limit on the rms
variation of the radius of the 20\% contour is 3\% (Paper III).  Since
our radio shell expansion velocity is that of a fit spherical shell
model and is therefore azimuthally averaged, the effect of the
occurrence of protrusion would be absorbed in the
statistical error of the expansion and does therefore not have to be
considered as a separate element in the error budget.  It needs
however to be considered again under item {4} below.

\item{3.} {\em Large scale anisotropy of the expansion of the optical
line emitting and absorbing ejecta:}

In view of the almost perfect circularity of the outer edge of the
radio shell it would be surprising if the geometry of the outer layers
of the ejecta were notably different. By contrast, the significant
linear polarization that was discovered in the early optical spectra
and continuum indicates that some asymmetries were likely present in
the ejecta up to a few weeks after shock breakout. These asymmetries
were successfully modelled with different geometries by H\"oflich
(1995), H\"oflich \etal\ (1996), and Tran \etal\ (1997) including one
with a spherical outer ejecta envelope and an off-center source
(H\"oflich 1995), such as may be expected since the progenitor of SN 1993J
was a member of a binary system (Maund \etal\ 2004, see also Podsiadlowski \etal\
1993).  Tran \etal\ (1997) also discussed a model where the ejecta
interacted with a clumpy and anisotropically distributed CSM.
 
In any case, in the plane of the sky, the ejecta envelope must have
been highly circular and expanding almost isotropically to be
consistent with the near circularity of the radio shell from a few
weeks (Bartel \etal\ 1994) to several years (Paper III) after shock
breakout.

What is the evidence for sphericity and almost isotropic expansion in
three dimensions?  First, the good match between the decrease of the
BVZI (abs, em) and the deceleration of the outer surface of the radio
shell.  Further, the almost exactly symmetrical and box-like shape of
the H$\alpha$ emission line after one year, as first noted by
Filippenko \etal\ (1994). In particular, for the 11 epochs from $t =
523$ to 2454~d, the mean of the ratios of the BVZI to RVZI values is
$-1.024$ and therefore the BVZI and RVZI values are almost equal in
magnitude, which is also reflected in the small differences of the
distance estimates of not more than 0.5 Mpc or 1.3\% relative to 3.96
Mpc (Table~\ref{t3dfit}, solutions 3, 4 relative to 1).  In fact the
late H$\alpha$ line profile, at $t\geq1000$~d, can be well fit by a
spherical shell of constant emissivity (Fransson \etal\ 2005) with a
shell thickness of 30\% of the shell's outer radius. We adopt an
uncertainty of our distance estimate due to these asymmetries of
0.05~Mpc or 1.3\%.

\item{4.} {\em Small scale anisotropy of the expansion of the optical
line absorbing and emitting ejecta:}

The development of Rayleigh-Taylor fingers at the contact surface
certainly leads to small scale anisotropies of the expansion of the
ejecta, but their contribution to the distance uncertainty is
minimal. First, Rayleigh-Taylor fingers likely change with time. However,
our distance estimates at early and late times are only different from
the estimate of 3.96 Mpc (solution 1 in Table~\ref{t3dfit}) by up to
0.05 Mpc or 1.3\%. Second, any anisotropy should be reflected in the
differences of the BVZI and RVZI\@. However, the distance solutions with
the H$\alpha$ BVZI em. and RVZI values are only different from the
estimate of 3.96~Mpc by 0.05 Mpc or again 1.3\%.  We
therefore conclude that any contribution to the distance error budget
due to small scale anisotropies is negligible or already included in
the contribution from the large scale anisotropies.

\item{5.} {\em The effect of possible biases of the angular velocity fits}

In Paper II we discussed five sources of possible systematic errors on
the determinations of the angular outer radius of the radio shell: i)
azimuthal modulation of the brightness along the ridge, ii) absorption
in the radio shell center, iii) thickness of the radio shell, iv)
radial modulation of the shell profile, and v) deviations of circular
symmetry of the shell. We estimated that the total resulting error of
the values of \thout\ is less than 5\% for the early epochs and
decreasing for the later epochs and largely included in the errors of
\thout. Here we estimate that the resulting error of \dotthout\ is
$<1.7$\% for $t=30$ to $\sim300$~d and negligible for later
epochs. The corresponding contribution to the error of the distance is
already largely included in the statistical contribution, since the
errors of the radio velocities extrapolated to the early times of the
H$\alpha$ abs. values are much larger than 1.7\%. Any remaining
contribution is almost certainly not larger than half the difference
between the distance solutions with H$\alpha$ BVZI abs. values at
early times and BVZI em. values at late times. We take an uncertainty
of our distance estimate due to these effects of 0.05~Mpc or 1.3\%.

\item{6.} {\em The possibility of a prolate or oblate geometry for the supernova}

We considered the possibility that the radio shell was in fact prolate
or oblate, but fortuitously aligned so that its projection in the
plane of the sky was almost circular, in other words aligned in such a
way that the long (prolate case) or short (oblate case) axis was
pointing toward us at a small angle.  The radial expansion velocity
would then be larger (prolate case) or smaller (oblate case) than the
transverse one, but the observed radial and transverse symmetry
properties would be retained.  Consider, for example, a prolate
spheroid, with the long axis 10\% larger than its other two axes.  A
numerical calculation shows that such a spheroid, randomly oriented,
has only a 7\% chance of being aligned so that its projection on the
plane of the sky is circular to within 1.4\%, as is observed for
SN~1993J (Paper I, see also Paper III).  In this case, we would
overestimate the distance by $\sim$9\% since the radial velocities
would in fact be larger than the tangential ones.  The error in the
distance and the odds {\em against}\/ fortuitous alignment are both
roughly proportional to the deviation of the long axis from the others.
The chances for the alignment of oblate objects are similarly small,
but the distance error would have the opposite sign.  Since such 
alignments are unlikely, we
do not consider them further.

\item{7.} {\em Spatial relation between the optical and radio
regions:}

The spatial relation between the radio and optical emission regions
which is conceptually shown in Figure~\ref{f2snmodel} is empirically
fairly well supported by the correspondence between our measurements
of the shell thickness and that predicted by analytical computations
(Chevalier \& Fransson 1994) and hydrodynamic simulations (Mioduszevski
\etal\ 2001). We think that the largest uncertainty of the spatial
relation is linked to the unknown growth of the Rayleigh-Taylor
fingers with time and the corresponding slight differences in the
deceleration of the H$\alpha$ line absorbing and emitting gas, the
contact surface and the outer surface of the radio shell or the
forward shock.  While the H$\alpha$ line absorbing gas (early times)
is less decelerated within 1$\sigma$, the H$\alpha$ line emitting gas
(late times) is less decelerated within 4$\sigma$ (solutions 2 and 3,
respectively, in Table~\ref{t2mfit}).  It appears that the H$\alpha$
emitting gas at its measured maximum velocity as given in
Table~\ref{t1vopt} is continuously expanding further into the 
space between the contact surface and the forward shock front. This is
a sign that Rayleigh-Taylor fingers are progressively eating further
into the shocked CSM.

The range of the possible extent of the Rayleigh-Taylor fingers
determines the distance error contribution due to this item, provided
that the gas in the Rayleigh-Taylor fingers remains in the range of the
temperature for Balmer line absorption and emission in the first place. 

We determined the error contribution through boundary conditions. The
upper limit was computed by assuming that at the earliest time the
hydrogen gas has not yet expanded into the shocked CSM through
Rayleigh-Taylor fingers but rather extends just to the contact
surface. We solved for the distance by fitting the angular velocity of
the contact surface to the BVZI abs. values from $t=7$ to $10$~d. We
get a value for the distance of $4.16\pm0.12$~Mpc.

The lower limit was computed by prohibiting the
hydrogen gas velocity from exceeding the forward shock velocity at any
time, since we have not yet seen any significant protrusions in the
radio shell images. 
Because of the deceleration of the hydrogen gas
being slightly weaker at late times than that of the forward shock,
to be conservative, the computation was done for the latest times. We solved for the
distance by fitting the angular velocity of the outer radio shell
surface to the H$\alpha$ and H$\beta$ BVZI em. values from $t=1766$ to
2454~d. We get a value for the distance of $3.67\pm0.08$~Mpc. 
We therefore adopt a (symmetric) error of 0.25 Mpc or 6.3\% as the
1$\sigma$ contribution to the distance error budget.

In case of the upper-limit distance of $D=4.16$~Mpc, the H$\alpha$
BVZI would not exceed the velocity of the contact surface for the
first $\sim$10~d and only marginally if at all exceed it for up to
$t\sim$700~d.  Only from then onward would the H$\alpha$ BVZI, and
from $t\sim2000$~d the H$\beta$ BVZI, significantly exceed the (mean)
contact surface velocity, indicative of the development of
Rayleigh-Taylor fingers into the shocked CSM.

In case of the lower-limit distance of $D=3.67$~Mpc, the H$\alpha$
absorption gas would expand with $97\pm6$\% of the velocity of the
forward shock already a few days after shock breakout, which would be
indicative of the Rayleigh-Taylor fingers having reached the forward
shock almost from the start. Also at later times, years after shock
breakout, the H$\alpha$ emission gas would keep expanding, within the
errors, with the velocity of the forward shock.  The H$\beta$ emission
gas would in fact have exceeded the velocity of the (mean) forward
shock by as much as 1.9$\sigma$, which is unlikely.

\end{trivlist}

\subsection{The distance solution with statistical and systematic errors combined.}
For the statistical standard error of the distance we take the
uncertainty from solution 1 in Table~\ref{t3dfit}.  For the systematic
errors we consider our elaborations above and the range of solutions
listed in Table~\ref{t3dfit}. The largest contribution is the
uncertainty of relating the locations of the optical and radio
emission, which is largely related to the uncertainty of the growth of
the Rayleigh-Taylor fingers over time.  The second largest
contribution comes from our estimates of, or limits on, anisotropies
of transverse expansion. The other contributions are related to
anisotropies of radial expansion and to changes of the distance
estimate as a function of time. They are minor in comparison.  We list
all contributions to the error budget in Table~\ref{t4err}. We add
these latter contributions to the statistical error in quadrature and
get a combined standard error of 0.29 Mpc or 7.3\%.  Our final value
of the distance to SN~1993J and its host galaxy M81 is therefore:

$$D=3.96\pm0.05 {\rm (stat.)}\pm0.29 {\rm (syst.)}~{\rm Mpc}$$ or
$$D=3.96\pm0.29~{\rm Mpc}.$$

\begin{figure}
\plotone{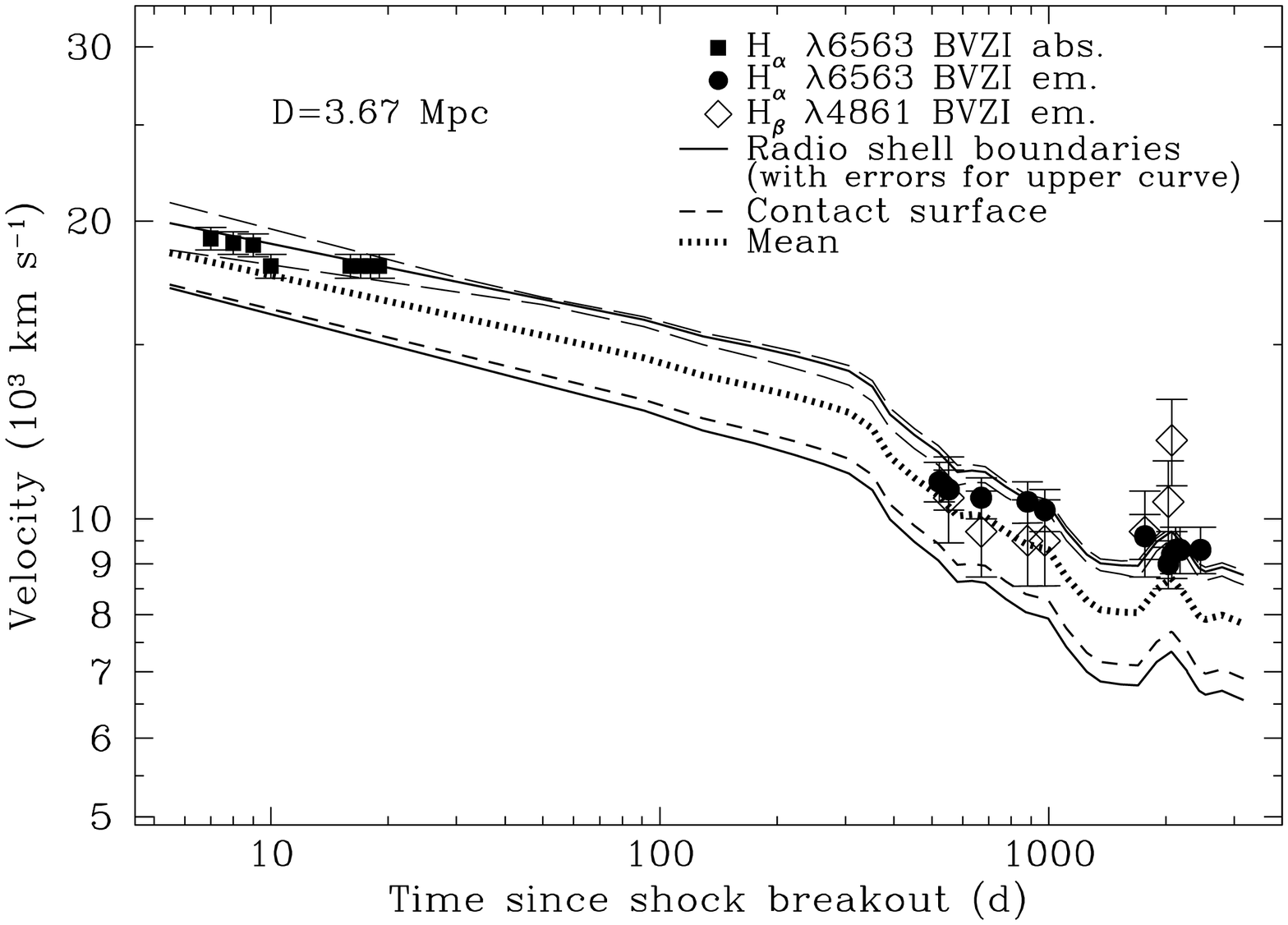}
\figcaption{As in Figure~\ref{f7halphablueradvellogmd}, but now the
velocity curves are plotted for a distance of 3.67 Mpc, which is the 
1$\sigma$ lower limit of our distance estimate of $3.96\pm0.29$~Mpc.
\label{f10halphahbetaradvelcdlogmd}}
\end{figure}

\begin{figure}
\plotone{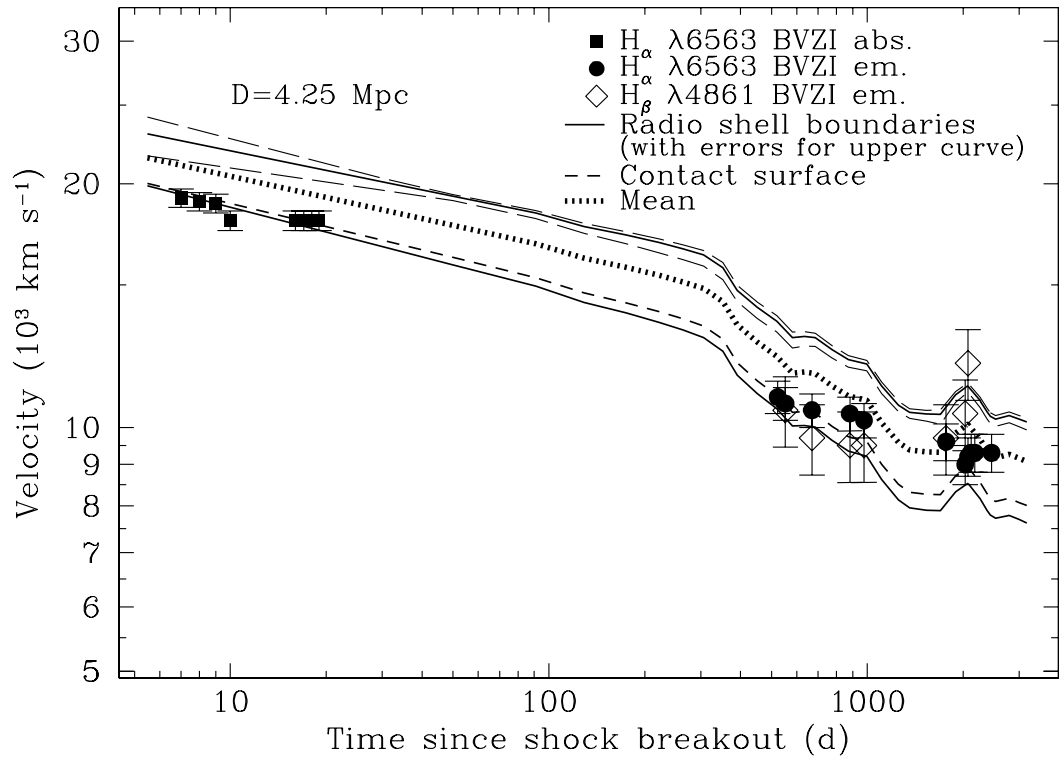}
\figcaption{As in Figure~\ref{f7halphablueradvellogmd}, but now the
velocity curves are plotted for a distance of 4.25 Mpc, which is the 
1$\sigma$ upper limit of our distance estimate of $3.96\pm0.29$~Mpc.
\label{f11halphahbetaradvelcdlogpd}}
\end{figure}

We plot the H$\alpha$ and H$\beta$ BVZI values with the radio velocity
curves for our upper (4.25 Mpc) and lower (3.67 Mpc) 1$\sigma$
distance limits in Figures~\ref{f10halphahbetaradvelcdlogmd} and
\ref{f11halphahbetaradvelcdlogpd}, respectively.

\section{DISCUSSION}
\label{disc}

The combination of our VLBI observations of SN 1993J with optical
spectral line observations from the time of explosion to several years
thereafter provides a unique opportunity to study the spatial and
dynamical relation between the radio shell and the optical line
emitting gas in the context of the circumstellar interaction between
the supernova ejecta and the surrounding hydrogen gas and to determine
the distance to the host galaxy M81 geometrically with the Expanding
Shock Front Method (ESM).

The multifrequency VLBI observations, phase-referenced to the core of
the host galaxy, have significantly advanced our knowledge of the
evolution of supernova radio shells. In our Galaxy, radio shells of
supernovae have been observed over at most $\sim10\%$ of their age.
SN 1993J has been observed essentially over 100\% of its age.  In the
first paper of this series, and of relevance to this discussion, we
reported the position of the explosion center with an accuracy of
about 160 AU in the galactic reference frame, and determined a 5.5\%
upper limit on any anisotropic expansion on the plane of the sky
(Paper I). In the second paper of the series we consistently
determined the rate of the expansion of the supernova from the
explosion center throughout most of the supernova's lifetime, and
determined changes of the rate, both with high accuracy (Paper II).
In the third paper we presented the series of images of the expanding
radio shell and investigated the structure changes and the emission
profile of the shell (Paper III). In this fourth paper we combine our
radio results on the speed of the expansion, the limit on anisotropy,
and on the emission profile of the radio shell, with computations of
the location of the contact surface, and optical observations of the
width and shape of spectral lines.

These results are important for discussions concerning 1) optical
line-emitting gas and its relation to the radio shell and
Rayleigh-Taylor instabilities, 2) possible misinterpretations of some
particular H$\alpha$ and He I lines reported in the literature, and 3)
the distance to the host galaxy M81 and the Hubble constant.  We will
discuss each of the aspects in turn.

\subsection{The optical line-emitting gas and its relation to the radio shell}
A few aspects of the results mentioned in the previous section deserve
further discussion.  First, the RVZI and RVHI H$\alpha$ values for
$t=16$~d to $\sim500$~d indicate a moderately sloped red side, with
the RVZI values well below the velocity of the reverse shock and with
the RVHI values showing no indication of deceleration (compare
Figures~\ref{f6halphaalllogmd} and
\ref{f7halphablueradvellogmd}). Second, for $t\gtrsim$500~d, the
H$\alpha$ profile is very symmetric, with steeply sloped blue and red
sides.  Third, the distance estimates with the H$\alpha$ abs. at early
times and H$\alpha$ em. at late times are equal to within 0.09 Mpc or
2\%.  And fourth, at late times the radio shell is slightly more
decelerated than the line emitting gas.

The first point may indicate that at early times the red side of the
H$\alpha$ line profile is affected significantly by the light passing
through the interior of the supernova and that with time and
decreasing density in the interior, this effect on the line profile
diminished. An alternative interpretation, that a couple of weeks
after shock breakout, the H$\alpha$ emitting gas was spread
irregularly through a much larger fraction of the ejecta on the far
side than on the near side of the expanding supernova, is less
attractive in view of the box-like emission profile seen later on. The
H$\alpha$ line emitting gas would have to develop from a one-sided
irregular geometry to a spherical shell with fairly well defined
boundaries, which is unlikely.

The second point indicates that the H$\alpha$ line emitting region is
very symmetric along the line of sight, at least for
$t\gtrsim$500~d. Indeed the high degree of symmetry, with the
magnitudes of the maximum velocities in the two radial directions
being equal to within 2.4\%, is similar to the degree of circular
symmetry in the plane of the sky, to within 1.4\% of the radio shell's
outer edge, making it highly likely that the radio shell as well as
the H$\alpha$ line emitting gas shell are spherically symmetric.

The third point indicates that, assuming that the maximum observed
velocities are indeed those of the tip of Rayleigh-Taylor fingers, any
evolution of such fingers from $t=7$~d to $\sim$3000~d is, on average,
rather constrained.  No matter whether we used a) \dotthout, b)
\dotthc, or c) \dotthx\ to solve for the distance, implicitly assuming
that these fingers a) reached the forward shock, b) were largely
constrained to the contact surface, or c) reached to half way in
between the contact surface and the forward shock, the distance
estimates for $t=523$ to 2454~d were larger by only a) 0.06$\pm$0.09,
b) 0.23$\pm$0.12, and c) 0.09$\pm$0.08~Mpc than those for $t=7$ to
19~d.  

The somewhat larger discrepancy of 0.23$\pm$0.12~Mpc when using
\dotthc, could have several interpretations. It could be taken as an
argument against the respective distance solution of 4.33~Mpc which is
anyway larger than our 1$\sigma$ upper limit of the distance.  It could
also be taken as information on the shell thickness at early times,
the evolution of the Rayleigh-Taylor fingers at late times, or the
separation of the contact surface from the forward shock, since using
\dotthc\ for the fit is, in comparison to using \dotthx\ or \dotthout\
more dependent on these parameters.  
The shell thickness would have to be slightly larger than assumed.
More precisely, for $t=7$ to 19~d, $r_{\rm fs}/r_{\rm rs}$ would have
to be 1.22$\pm$0.03, rather than 1.16 as assumed, in order to give a
better fit, that is the same distance for $t=7$ to 19~d as for $t=523$
to 2454~d.  As an alternative, the Rayleigh-Taylor fingers would have
to have only slightly expanded into the shocked CSM toward the forward
shock at $t=\sim1000$ to 3000~d. (see
Figure~\ref{f10halphahbetaradvelcdlogmd}). Lastly, it is possible that
at late times the contact surface itself or the base of the
Rayleigh-Taylor fingers has moved closer to the forward shock than
assumed. Taking into consideration any of these alternatives would
improve the quality of the fit to that obtained using \dotthx\ or
\dotthout. In other words, there is in fact marginal evidence that one
or more of these alternatives actually apply.

The fourth point indicates that 
the comparison of the radio and optical deceleration may provide a
sensitive way of probing the evolution of the Rayleigh-Taylor fingers
for particular time intervals. While the radio and optical
decelerations for the total time and at early times are equal within
the combined standard errors, the radio deceleration is slightly but
significantly larger than the optical deceleration for
$t\geq520$~d. This difference may indeed be indicative of the
Rayleigh-Taylor fingers eating into the shocked CSM\@.  The O[III] and
Na I line emitting gas is also significantly decelerated, slightly
less than the H$\alpha$ line emitting gas, but within the errors of
the former. Its velocity is 12\% smaller and may not be as much
influenced by the reverse shock as the H$\alpha$ gas. A smaller
deceleration, if any at all, is therefore expected.

\subsection{The distance to M81}

The nearby spiral galaxy M81 is considered an important one for the
determination of the extragalactic distance scale and the Hubble
constant, $H_0$, since it is used as a calibrator for various methods
of determining distances and also since it can be used for a
comparison of the various methods. How does our ESM distance compare
with other distance determinations to M81 or SN~1993J? In
Table~\ref{t5distdet} we compare our determination with those obtained
with methods that depend on an absolute calibration scheme and those
obtained with methods that, like ESM, give the distance directly.
These latter methods are the Expanding Photosphere Method, or EPM
(Kirshner \& Kwan 1974; Eastman \& Kirshner 1989; Eastman, Schmidt, \&
Kirshner 1996) which is a variation of the Baade-Wesselink method
first suggested by Baade (1926) for variable stars and later modified
and applied by Wesselink (1946), and the similar Spectral-Fitting
Expanding Atmosphere Method, or SEAM (Baron \etal\ 1995). Both methods
derive the distance by assuming spherical symmetry and, in effect,
combining an estimate of the angular radius with the linear radius of
the photosphere or the line-forming region. The former further assumes
that the spectrum of the supernova is approximately that of a
blackbody and computes the angular radius via spectral photometry and
the linear radius from the velocity of spectral lines and an estimated
time of shock breakout. The latter uses non-LTE radiative transfer
codes and fits observed spectra with synthetic spectra to determine
the spectral energy distribution and the angular and linear radii of
the supernova at any given time.

The difficulty in deriving distances with EPM is that a velocity needs
to be derived from the Doppler shift of those lines that are assumed
to best match the photospheric velocity, and that a correction factor
needs to be computed to account for the differences between the
supernova spectral energy distribution and that of a
blackbody. Recently, Dessart \& Hillier (2005) reported on new
computations of these correction factors that would increase the EPM
distances by Eastman \etal\ (1996) by 10 to 20\%. A direct comparison
between the EPM distance for SN~1999em of $7.5\pm0.5$~Mpc (Hamuy
\etal\ 2001), $8.2\pm0.6$~Mpc (Leonard \etal\ 2002), and
$\sim7.83$~Mpc (Elmhamdi \etal\ 2003), with the Cepheid distance for
the host galaxy, NGC 1637, of $11.7\pm1.0$~Mpc (Leonard \etal\ 2003)
shows the range of discrepancies.

The difficulty with SEAM is to compute radiation transport in a
rapidly expanding supernova atmosphere so that the resulting synthetic
spectra accurately match the observed ones. For example, the distance
to SN~1999em derived with SEAM was reported to be $12.5\pm1.8$~Mpc
(Baron et al. 2004), 7\% larger than, but within the errors of, the
Cepheid distance.

In comparison to at least EPM, ESM has in principle more observable
parameters and fewer systematic uncertainties. In ESM, the transverse
expansion velocity is directly measured and not dependent on dilution
factors.  Changes of the expansion velocity can be measured
directly. The distance determination is almost completely independent
of the assumed date of shock breakout. The isotropy of the expansion
can also be measured directly, at least in projection. What is similar
to EPM and SEAM is the difficulty of relating the Doppler shift of the
spectral lines to the transverse expansion velocity.

Our small statistical uncertainty of only 1.3\% reflects the good fit
of the VLBI expansion curve to the maximum Doppler shift of the
H$\alpha$ line. Our larger systematic uncertainty of $7.2$\% includes
a 6.3\% standard error due to the uncertainty of combining optical and
radio velocities, and, omitted from published EPM or SEAM distance
determinations, a 3\% uncertainty from our measurements on the
anisotropy of the transverse expansion of the shock fronts.

Our value of $3.96\pm0.29$~Mpc is in the upper half of previously
determined distances with ESM and SEAM, but consistent with them.  The
Cepheid distance estimate of $3.93\pm0.26$~Mpc by Huterer \etal\
(1995) is virtually the same as our ESM distance estimate.  In this
context it is worth noting that recent distance estimates for the
galaxy M82 and the dwarf elliptical galaxies, F8D1 and BK5N have
also been similar to our distance estimate for M81. 
The galaxies are all in the same group, and a HI map shows
evidence that M81 and M82 have tidally interacted recently. They have
a projected separation of only $\sim40$ kpc. Using the tip of the red
giant branch method, Madore reported a distance of $3.9\pm0.4$~Mpc for
M82 and Caldwell \etal\ (1998) distances of 3.98$\pm$0.15~Mpc for F8D1
and $3.80\pm0.27$~Mpc for BK5N. These distance estimates are in good
agreement with our M81 distance estimate.

In contrast,  our distance estimate is
9$\pm$13\% larger than the Cepheid distance estimate of
$3.63\pm0.34$~Mpc by Freedman \etal\ (1994), In other words, if the distance to M81 were indeed
3.63 Mpc, which is only slightly below our $1\sigma$ lower limit of
3.67 Mpc, then the H$\alpha$ absorption between $t=7$ and $19$~d and
the later H$\alpha$ and H$\beta$ emission at $t\sim2000$ would have to
occur right up to the forward shock (see
Figure~\ref{f10halphahbetaradvelcdlogmd}).  The most likely
interpretation for this scenario would be that the Rayleigh-Taylor
fingers developed strongly right after shock breakout, reached the
forward shock at $t\sim10$~d and and continued to reach the forward
shock up to at least $t\sim2500$~d. We consider this scenario to be
less likely and instead argue in favor of a larger distance.

Direct and Cepheid distance determinations can also be compared for
the nearby galaxies M33 and NGC 4258. Using a detached eclipsing
binary, the distance to M33 was found to be 0.96$\pm$0.05~Mpc (Bonanos
et al. 2006), 14\% or $1.5\times$ the combined uncertainty larger than
the Cepheid distance of 0.84$\pm$0.06~Mpc (Freedman et al. 2001). In
contrast, the distance derived with maser VLBI is
0.73$^{+0.17}_{-0.14}$~Mpc (Brunthaler et al. 2005), 13\% or
0.6$\times$ the combined uncertainty smaller than the Cepheid
distance.  For NGC 4258, using maser VLBI, the distance is
$7.2\pm0.5$~Mpc (Herrnstein \etal\ 1999). This value is 10\% or
1$\times$ the combined uncertainty smaller than the Cepheid distance
of $7.98$~Mpc (systematic uncertainty not given, total error assumed
to be 0.6~Mpc, i.e., the same as that given by Newman \etal\ 2000;
compare with 8.1$\pm$0.4~Mpc by Maoz \etal\ 1999 and $7.8\pm0.6$~Mpc
by Newman \etal\ 2000).  It appears that, at least for our examples,
distances derived with ESM, SEAM, and by using a detached eclipsing
binary, are larger than the Cepheid distances by 7 to 14\% while the
distances derived with maser VLBI tend to be somewhat smaller.
Perhaps the systematic uncertainties are still underestimated.

\subsection{The Hubble Constant}

Is it possible to derive $H_0$ from our distance estimate alone?
Recently it has been reported that the Hubble flow can be extrapolated
back to about 2 Mpc if it is corrected for peculiar motions caused by
the large-scale distribution of matter in the nearby universe (e.g.,
Karachentsev \etal\ 2003; Sandage \etal\ 2006). Several authors give
corrected flow velocities for M81 or the M81 group: Tonry \etal\
(2000) give 246 \kms\ (given in Freedman \etal\ 2001 for M81) and
Sandage \etal\ (2006) give 234 \kms. With our distance estimate, these
flow velocities would lead to $H_0$ of $62\pm5$ and $59\pm4$ \kmsMpc,
respectively.  Karachentsev \etal\ (2002) give a low velocity, of 106
\kms, but a rather large velocity of 360 \kms\ for the spatially
nearby galaxy M82.  Freedman \etal\ (2001) give a similarly low
velocity for M81, of 80 \kms.  Such low velocities would lead to
unrealistically small values of $H_0$.  Clearly, the range of
predicted flow velocities for M81 is still too large to be used for a
reliable estimate of $H_0$ and our distance estimate for M81 can at
best be taken as an indication against a relatively large value for
$H_0$.

Is it possible to derive $H_0$ by anchoring the Cepheid distance scale
to our distance estimate for M81? In this respect the Cepheid distance
modulus of M81 relative to that of the Large Magellanic Cloud (LMC)
may be of interest since the LMC is in many studies crucial for the
entire Cepheid distance scale.  However, different authors derive
different relative distance moduli.  For instance, Freedman \etal's
(1994) distance modulus for M81 of $27.80\pm0.20$~mag, corresponding to a 
distance, $D_{\rm M81}=3.63\pm0.34$~Mpc, is based on the
assumption of an LMC modulus of 18.50 mag ($D_{\rm LMC}=50.1$~kpc).

In contrast, Huterer \etal\ (1995) obtained, with a different analysis
of the same HST data, an M81 modulus of $27.97\pm0.14$~mag ($D_{\rm
M81}=3.93\pm0.26$ Mpc).  This estimate, as mentioned earlier, is essentially identical
to our estimate for the distance of M81. In the same analysis they
also obtained an LMC modulus of $18.45\pm0.10$ mag ($D_{\rm
LMC}=49.0\pm2.3$ kpc).  Apparently, different analyses led to
distance ratios different by 10\%.

Huterer \etal's (1995) LMC distance modulus can be compared with that
of $18.41\pm0.16$ mag obtained by Macri \etal\ (2006) through a
determination of the Cepheid distance modulus of the maser-host galaxy
NGC 4258 relative to that of the LMC and who give an estimate of
$H_0=74\pm3 {\rm (stat.)} \pm6 {\rm (syst.)}$~\kmsMpc.  It can also be
compared with $18.39\pm0.05$ mag obtained through revised Hipparcos
parallaxes of Cepheids by van Leeuwen \etal\ (2007) who also argue in
favor of a revision of Sandage \etal's (2006) and Freedman \etal's
(2001) values for the Hubble constant to $H_0=70\pm5$~\kmsMpc\ and
$H_0=76\pm8$~\kmsMpc, respectively.

Which value for $H_0$ can we derive with our distance estimate for M81
in view of this discussion?  Freedman \etal\ (2001) obtained, on the
basis of their Cepheid analysis, $H_0=72\pm8$~\kmsMpc. If the
difference between their M81 distance determination and ours is due to
a systematic effect, then, on the basis of their Cepheid analysis,
scaling would lead to $H_0=66\pm11$~\kmsMpc, where the uncertainty is
one standard error, derived by adding errors in quadrature. This value
can be compared with $H_0=62.3\pm$1.3(stat.)$\pm$5.0(syst.)~\kmsMpc\
(Sandage \& Tammann 2006). Huterer \etal's (1995) Cepheid
distance determinations for M81 and the LMC, however, are consistent
with a $\sim10$\% higher value for $H_0$.

In conclusion, our distance estimate for M81, which is somewhat larger than
Freedman \etal's (1994) estimate, may argue for a correspondingly smaller value
of $H_0$. However, deriving a reliable value of $H_0$ from our M81 distance
determination directly would require a more reliable estimate of M81's flow
velocity than is presently available. Deriving such value of $H_0$ indirectly
would require anchoring other distance scales more reliably to our M81 distance
determination.

\section{CONCLUSIONS}
\label{conc}

\noindent Here we list a summary of our main conclusions.

\begin{trivlist}

\item{1.} SN 19993J is the best suited supernova yet for determining
an accurate geometric distance with the expanding shock front method,
ESM, by comparing velocities from the Doppler shift of optical lines
with the VLBI determined velocities of the radio shell.

\item{2.} The brightness profile of the radio shell being consistent
with a sharp intensity decline at the outer rim argues in
favor of the outer surface of the radio shell being indeed equivalent
to the expanding forward shock front.  The thickness of
the radio shell being consistent with predictions from hydrodynamic
simulations argues in favor of the inner surface of the radio shell
being equivalent to the reverse shock front.

\item{3.} The angular expansion velocity of the outer surface of the radio
shell could be determined for the time from $t\sim30$ to
$\gtrsim3000$~d with an uncertainty at early times of $\lesssim$5\%
and at late times of $\lesssim$2\%.

\item{4.} The average of the thickness of the radio shell, which is 29\%
of the radius of the outer radio shell surface for the time
$t\sim1000$ to 3000~d, coincides with the average of the thickness 
predicted by hydrodynamic simulations for that time interval.

\item{5.} Changes of the velocity of the radio shell's outer
surface at $t\sim350$ and $2000$~d, which are reflected by sharp
minima in the X-ray lightcurve at these same times, have only partial
correspondence in the velocities from the optical lines. At
$t\sim350$~d, only H$\alpha$ RVZI and RVHI were recorded and do not
show any correspondence, providing evidence that the observed
H$\alpha$ velocities at the far side of the expanding supernova were
biased by absorption and do not reflect the velocity of the contact
surface.  At the less dominant change at $t\sim2000$~d, only the
velocities from the H$\beta$ and O[III] lines appear to be influenced.

\item{6.} The H$\alpha$ gas has with $m=0.868\pm0.004$ virtually the
same (mean) deceleration as the outer surface of the radio shell (forward
shock, $m=0.870\pm0.005$) and the surface midpoint between the outer surface
and the contact surface ($m=0.864\pm0.008$) for the total time between
$t=7$ and $\sim$2500~d.  The decelerations are also equal within the
errors for the early time interval between $t=7$ and $\sim$300~d.  For
the late time interval from $t\sim 500$ to $\sim$2500~d the H$\alpha$
gas is with $m=0.890\pm0.022$ slightly less decelerated than any of
the surfaces of the radio shell.

\item{7.} It is possible that the difference in the deceleration is
due to the development of Rayleigh-Taylor fingers.

\item{8.} The O[III] and Na I gas is also decelerated, although only
with marginal significance.

\item{9.} Using ESM by combining the BVZI values of the H$\alpha$ line
with the angular velocity half way between that of the contact surface
and that of the radio shell's outer surface gives a geometric distance
estimate of $D=3.96\pm0.29$~Mpc. The standard error combines in
quadrature a statistical contribution of 0.05~Mpc and a systematic
contribution of 0.29~Mpc.

\item{10.} The distance estimate depends by not more than 0.06 Mpc or 1.5\% on
whether the estimate is obtained with only a) early H$\alpha$ BVZI data from $t=7$ to
19~d, b) late H$\alpha$ BVZI data from $t\sim500$ to $\sim2500$~d, c) H$\alpha$ RVZI
data, or d) H$\beta$ BVZI data.

\item{11.} The largest contribution to the error comes from the
uncertainty of relating the optical velocities to the radio
velocities.

\item{12.} The $1\sigma$ upper limit of 4.25 Mpc corresponds to the H$\alpha$
absorption occurring at the earliest time ($t\sim 10$~d) only up to the reverse shock. 
The $1\sigma$ lower limit of 3.67 Mpc corresponds to the H$\alpha$
absorption or emission and  H$\beta$ emission occurring as far out as the 
forward shock which would
mean that Rayleigh-Taylor fingers have developed and stretched out to
the forward shock within the errors, from essentially $t\sim 10$ to $\sim 2000$~d. 

\item{13.} The distance estimate is $9\pm13$\%, or 0.7 $\times$ the combined
errors, higher than the Cepheid distance of 3.63$\pm$0.34~Mpc by
Freedman \etal\ (1994).

\item{14.} If the distance to M81 were indeed 3.63 Mpc, then the
H$\alpha$ absorption at $t\sim10$~d and the H$\alpha$ and H$\beta$
emission at $t\sim 2000$~d would have to originate as far out as the
forward shock. The most reasonable interpretation would be that
Rayleigh-Taylor fingers had developed right after shock breakout and
stretched out as far as the forward shock. We think that this scenario is
less likely and that the distance has to be larger.

\item{15.}  Our direct distance estimate for M81 may argue for a value of
$H_0$ somewhat smaller than that of Freedman \etal\ (2001).  However,
to derive a more reliable value of $H_0$ from our M81 distance
determination, a more reliable Hubble flow velocity estimate or a reliable
way of anchoring other distance scales to the distance of M81 would be needed. 

\end{trivlist}

\acknowledgements
\noindent
Acknowledgements: 
We thank V. I. Altunin, A. J. Beasley, W. H. Cannon,
J. E. Conway, D. A. Graham, D. L. Jones, A. Rius, G. Umana, and
T. Venturi for help with several aspects of the project.  J. Cadieux,
M. Craig, M. Keleman, and B. Sorathia helped with some aspects of the VLBI data
reduction during their tenure as students at York. N. B. thanks the
Canadian Institute for Theoretical Astrophysics (CITA), Toronto,
Canada, Observatorio do Valongo, Rio de Janeiro, Brazil, and the
Harvard-Smithsonian Center for Astrophysics, Cambridge, U.S.A. for
support during part of his sabbatical leave from York University in
which this research was done.  V. V. D. would like to
acknowledge useful discussions with, and comments from, Paolo Mazzali,
Claes Fransson, Roger Chevalier, and Amy Mioduszewski. V. V. D.'s
research is supported by NSF grant AST 0319261 to the University of
Chicago and by NASA through grant HST-AR-10649 from the Space
Telescope Science Institute, which is operated by AURA under NASA
contract NAS5-26555.  We thank NRAO, The European VLBI Network, the
NASA/JPL Deep Space Network (DSN), and Natural Resources Canada for
providing support for the observations. Research at York University
was partly supported by NSERC\@.  NRAO is operated under license by
Associated Universities, Inc., under cooperative agreement with NSF\@.
The NASA/JPL DSN is operated by JPL/Caltech, under contract with NASA.
We have made use of NASA's Astrophysics Data System Abstract Service.
\clearpage

\appendix

\section{Possible misinterpretations of some particular H$\alpha$ and He I 
lines by others}\label{comparison}
 
Matheson \etal\ (2000b) reported a couple of H$\alpha$ BVZI values of
16,600 and 16,100 \kms, at $t=433$ and 473~d, respectively, and He I
$\lambda$5876 BVZI values of $\sim$16,000 \kms between $t= 553$ and
2454~d that appear puzzling in the context of the other optical
velocities and the radio shell velocities.  Each of the values is
clearly larger than the velocity of the outer surface of the radio
shell, at late times even twice as large. The couple of H$\alpha$ BVZI
values are almost certainly overestimates due to blending of the
H$\alpha$ line with the O[I] line. On first sight it is intriguing
that the apparent large decrease in the H$\alpha$ velocity from 16,600
and 16,100 \kms\ at $t=433$ and 473~d to 10,400 and 10,000~\kms\ at
$t=523$ and 533~d, respectively, approximately coincides with the
large deceleration of the radio shell which in turn coincides with an
increase in the X-ray flux (Zimmermann \& Aschenbach, 2003).  However,
we think that the diminishing effect of the blending of the H$\alpha$
line at the time of the large radio deceleration is coincidental. The
He I values were derived from a decomposition of the line
profile. This decomposition may be questionable in light of the values
being so large. We did not list these H$\alpha$ and He I velocity
values in Table~\ref{t1vopt}.

\clearpage

\vfill\eject

\begin{deluxetable}{c r r c r c c r c}
\tablecaption{Expansion velocities of SN~1993J from optical lines \label{t1vopt}}
\tablewidth{0pt}
\tablehead{ 
\colhead{Age\tablenotemark{a}} & 
\colhead{H$\alpha$ $\lambda6563$} & \colhead{H$\alpha$ $\lambda6563$} &
\colhead{H$\alpha$ $\lambda6563$} & \colhead{H$\alpha$ $\lambda6563$} &
\colhead{H$\beta$ $\lambda4861$} & \colhead{[OIII] $\lambda5007$} &
\colhead{Na I $\lambda5890$} & \colhead{Reference\tablenotemark{b}} \\
  & 
\colhead{BVZI\tablenotemark{c}} & \colhead{BVHI\tablenotemark{c}} & \colhead{RVZI\tablenotemark{c}} & 
\colhead{RVHI\tablenotemark{c}} & \colhead{BVZI\tablenotemark{d}} & \colhead{RVZI\tablenotemark{d}} & 
\colhead{RVZI\tablenotemark{d}} & \\
\colhead{(d)} & 
\colhead{(\kms)} & \colhead{(\kms)} & \colhead{(\kms)} & \colhead{(\kms)} & \colhead{(\kms)} & \colhead{(\kms)} &
\colhead{(\kms)} }
\startdata
7    & $-19,200\pm$500\tablenotemark{e}&                &                 & & & & & 1 \\
8    & $-19,000\pm$500\tablenotemark{e}&                &                 & & & & & 1 \\
9    & $-18,900\pm$500\tablenotemark{e}&                &                 & & & & & 1 \\
10   & $-18,000\pm$500\tablenotemark{f} &                &                 & & & & & 1, 2, 4 \\
16   & $-18,000\pm$500\tablenotemark{f} &                &  14,000$\pm$500 & 6,000$\pm$500 & & & & 2, 4 \\
17   & $-18,000\pm$500\tablenotemark{f} &                &  14,000$\pm$500 & 6,000$\pm$500& & &  & 2, 4 \\
18   & $-18,000\pm$500\tablenotemark{f} &                &  14,000$\pm$500 & 6,000$\pm$500& & &  & 2, 4 \\
19   & $-18,000\pm$500\tablenotemark{f} &                &  14,000$\pm$500 & 6,000$\pm$500& & &  & 3, 4 \\
     &                &                &                 & & & &      &   \\
123  &                &                &  12,600$\pm$500 & 5,500$\pm$500 &                  &              &                  & 4 \\
171  &                &                &  11,420$\pm$400\tablenotemark{g} &               &                  &              &                  & 5 \\
182  &                &                &  11,300$\pm$500\tablenotemark{h}&               &                  &              &                  & 4 \\
236  &                &                &  11,570$\pm$400\tablenotemark{g} &               &                  &              &                  & 5 \\     
255  &                &                &  11,370$\pm$400\tablenotemark{g} &               &                  &              &                  & 5 \\
286  &                &                &  11,200$\pm$500\tablenotemark{g} & 7,600$\pm$500 &                  &              &                  & 4 \\
298  &                &                &  11,500$\pm$500 & 7,400$\pm$500 &                  &              &                  & 4 \\
299  &                &                &  11,450$\pm$400\tablenotemark{g} &               &                  &              &                  & 5 \\
315  &                &                &  11,500$\pm$500 & 7,300$\pm$500 &                  &              &                  & 4 \\
367  &                &                &  11,460$\pm$400\tablenotemark{g} &               &                  &              &                  & 5 \\
355  &                &                &  11,500$\pm$500 & 7,600$\pm$500 &                  &              &                  & 4 \\ 
387  &                &                &  11,500$\pm$500 & 8,200$\pm$500 &                  &              &                  & 4 \\
433  &                &                &  11,500$\pm$500 & 8,200$\pm$500 &                  &              &                  & 4 \\
473  &                &                &  10,700$\pm$500 & 8,800$\pm$500 &                  &              &                  & 4 \\
523  & $-10,900\pm$500 & $-9,100\pm$500 &  10,400$\pm$500 & 8,400$\pm$500 &                  &              &                  & 4 \\
553  & $-10,700\pm$500 & $9,100\pm$500 &  10,500$\pm$500 & 8,200$\pm$500 & $-10,500\pm$1,050      & \phn9,100$\pm$910   & 8,400$\pm$840   & 4 \\
670  & $-10,500\pm$500 & $-8,800\pm$500 &  10,500$\pm$500 & 8,200$\pm$500 & \phn $-9,700\pm$  970 & \phn9,600$\pm$960   & 9,200$\pm$920   & 4 \\ 
881  & $-10,400\pm$500 & $-8,600\pm$500 &  10,000$\pm$500 & 8,200$\pm$500 & \phn $-9,500\pm$  950 &    10,000$\pm$1,000 & 8,700$\pm$870   & 4 \\
976  & $-10,200\pm$500 & $-8,700\pm$500 &   9,600$\pm$500 & 7,900$\pm$500 & \phn $-9,500\pm$  950 & \phn8,000$\pm$800   & 8,900$\pm$890   & 4 \\
1766 & $-9,600\pm$500 & $-8,300\pm$500 &   9,300$\pm$500 & 7,700$\pm$500 &  \phn $-9,700\pm$ 970  & \phn7,600$\pm$1,520 & 8,500$\pm$850   & 4 \\
2028 & $-9,000\pm$500 & $-7,600\pm$500 &   9,500$\pm$500 & 7,300$\pm$500 &      $-10,400\pm$1,040 & \phn9,800$\pm$980   & 8,400$\pm$840   & 4 \\
2069 & $-9,200\pm$500 & $-7,600\pm$500 &   9,100$\pm$500 & 7,300$\pm$500 &      $-12,000\pm$1,200 & \phn7,500$\pm$870   & 8,700$\pm$870   & 4 \\
2115 & $-9,300\pm$500 & $-7,600\pm$500 &   9,000$\pm$500 & 7,300$\pm$500 &      ($-14,900\pm$2,980) & \phn7,200$\pm$1,440 & 8,300$\pm$830   & 4 \\
2176 &  $-9,300\pm$500 & $-7,800\pm$500 &   8,700$\pm$500 & 7,400$\pm$500 &    ($ -11,800\pm$2,360) & \phn6,700$\pm$1,340 & 7,700$\pm$770   & 4 \\
2454 &  $-9,300\pm$500 & $-8,200\pm$500 &   9,200$\pm$500 & 7,200$\pm$500 &    ($ -13,400\pm$2,680) & \phn8,400$\pm$1,680 & 7,500$\pm$750   & 4 \\

\enddata 
\tablenotetext{a}{\scriptsize Time since  assumed  explosion date of 1993
Mar. 27.5 UT\@. Note, that this time is 0.5 d earlier than the time we
assumed in general throughout this paper and Papers I, II, and
III\@. The difference is not significant in any of our computations.}
\tablenotetext{b}{\scriptsize (1) Finn \etal\ 1995; (2) Lewis \etal\ 1994; 
(3) Trammell \etal\ 1993;  (4) Matheson \etal\ 2000a; (5) Patat \etal\ 1995.}
\tablenotetext{c}{\scriptsize BVZI: maximum velocity on the blue edge of the line;
BVHI: velocity at half of the maximum
intensity on the blue side of the line; RVZI: maximum velocity at zero
intensity on the red edge of the line; RVHI: velocity on the red side
of the line at half of the maximum intensity. The lines are emission
lines except on days 7 to 19, see (d, e). We determined the velocity
values from the published line profiles, except where indicated
otherwise, see (f).  The uncertainties are estimated standard errors,
given by the uncertainty of determining the baseline of the profiles,
the values at the baseline, the maximum of the profile and the values
at the half maximum. For more information, see text.}
\tablenotetext{d}{\scriptsize The velocity values and their errors were 
taken as listed in the literature. The values in parentheses have extra large errors and were not considered further in our analysis.}
\tablenotetext{e}{\scriptsize Maximum velocities obtained by adding 1400 \kms\ to
the reported velocities for the minimum of the absorption trough of
17,800, 17,600, and 16,800 \kms\ (Finn \etal\ 1995). The offset of
1400 \kms\ was obtained from the difference of the reported velocity
for the minima of the troughs of 16,600 \kms\ (Finn \etal\ 1995) and
the measured maximum velocity of 18,000 \kms\ from a published line
profile (Lewis \etal\ 1994).}
\tablenotetext{f}{\scriptsize Maximum expansion velocity from the blue 
edge of the absorption trough of the H$\alpha$ line.}
\tablenotetext{g}{\scriptsize Values listed by Patat \etal\ 1995. Profiles were not 
plotted for all these epochs.}
\tablenotetext{h}{\scriptsize Only the RVZI value is given. The RVHI value is 
omitted because blending of the O[I] line made identifying the 
maximum of the H$\alpha$ profile uncertain by a large margin.}
\end{deluxetable}

\begin{deluxetable}{c c c c c c c c c}
\tabletypesize{\scriptsize}
\tablecaption{The deceleration parameter for the radio and optical data\label{t2mfit}}
\tablewidth{0pt}
\tablehead{ 
\colhead{\#} & \colhead{Radio range} &
\multicolumn{2}{c}{Optical range\tablenotemark{a}} & \colhead{Lines\tablenotemark{a}}  & 
\multicolumn{3}{c}{Radio decel.\tablenotemark{b}} & \colhead{Optical decel.\tablenotemark{b}} \\
& \colhead{$t$(d)} &  \colhead{$t$(d)}  & \colhead{$v$(\kms)} & 
  & \colhead{$m_{\rm o}$} & \colhead{$m_{\rm c}$} & \colhead{$m_{\rm x}$ }   & \colhead{$m_{\rm opt}$} }
\startdata

1 & \phn30 -- 2,432 & \phn7 -- 2,454 & 19,200 -- 9,000 & H$\alpha$ abs., em. (BVZI) & $0.870\pm0.005$ & $0.855\pm0.008$ & $0.864\pm0.008$   &  $0.867\pm0.004$  \\
2 & \phn30 -- 306    & \phn7  -- \phn\phn19 & 19,200 -- 18,000 & H$\alpha$ abs. (BVZI)& $0.919\pm0.019$ & $0.904\pm0.020$ & $0.912\pm0.020$ & $0.936\pm0.023$  \\
3 & \phn520 -- 2,432 & \phn523 -- 2,454& 10,900 -- 9,000    & H$\alpha$ em. (BVZI)   & $0.798\pm0.007$ & $0.782\pm0.009$ & $0.791\pm0.009$ &  $0.890\pm0.022$  \\
4 & \phn520 -- 2,432 & \phn523 -- 2,454& 10,700 -- 8,700    & H$\alpha$ em. (RVZI)   & $0.798\pm0.007$ & $0.782\pm0.009$ & $0.791\pm0.009$ &  $0.897\pm0.023$  \\
5 & \phn520 -- 2,432 & \phn553 -- 2,454& 12,000 -- 9,500    & H$\beta$  em. (BVZI)   & $0.798\pm0.007$ & $0.782\pm0.009$ & $0.791\pm0.009$ &  $1.062\pm0.063$  \\
6 & \phn30 -- 2,432  & \phn7 -- 2,454 & 19,200 -- 9,000    & all of \# 1 - 5         & $0.870\pm0.005$ & $0.855\pm0.008$ & $0.864\pm0.008$ &  $0.865\pm0.003$  \\
  &                 &                 &                     &                        &                 &                 &                 &                   \\
7 & \phn520 -- 2,432 & \phn553 -- 2,069& 10,000 -- 7,500    & O[III] em. (RVZI)      & $0.798\pm0.007$ & $0.782\pm0.009$ & $0.791\pm0.009$ &  $0.908\pm0.066$  \\
8 & \phn520 -- 2,432 & \phn553 -- 2,454&  9,200 -- 7,500    & Na I em. (RVZI)        & $0.798\pm0.007$ & $0.782\pm0.009$ & $0.791\pm0.009$ &  $0.929\pm0.048$  \\

\enddata

\tablenotetext{a}{As in Table~\ref{t1vopt}.}  
\tablenotetext{b}{The
deceleration parameter, $m$, with $r\propto t^m$ and $v\propto
mt^{m-1}$, for the outer radius of the radio shell ($m_{\rm o}$), the
radius of the contact surface ($m_{\rm c}$), the radius midpoint
between them ($m_{\rm x}$), and the velocity of the ejecta gas with
optical line absorption and emission ($m_{\rm opt}$). See
\S~\ref{radoptdecel} for more information. The solutions of $m_{\rm
o}$ have a Chi-square per degree of freedom, $\chi_{\nu}^{2}$, close
to unity. The errors are adjusted for $\chi_{\nu}^{2}=1$.
The errors in $m_{\rm c}$ and $m_{\rm x}$ are taken from those of
$m_{\rm o}$ added in quadrature to 0.005 assumed to be the error from
the theoretical models. The solutions of $m_{\rm opt}$ have
$\chi_{\nu}^{2}$ between 0.1 and 0.7 for the hydrogen gas, 1.5 for the
O[III] gas and 0.3 for the Na I gas. The errors are not scaled for
$\chi_{\nu}^{2}=1$, since we think that our measurements of the
velocities from the line profiles are somewhat correlated. For
solution 1 and 6, we used data on $t=30$~d and between $t=520$ and
2,432~d to best match the time range in which optical data were used.
The solution was forced to fit the data point at $t=30$~d exactly to
give appropriate weight to this early point.}
\end{deluxetable}

\begin{deluxetable}{c l l l c c c}
\tablecaption{Distance estimates from combined radio and optical velocity data\label{t3dfit}}
\tablewidth{0pt}
\tablehead{
\colhead{Solution} & 
\multicolumn{2}{c}{Optical data range\tablenotemark{a}}& \colhead{Lines\tablenotemark{b}}  & 
\colhead{Fit distance\tablenotemark{c}} & \colhead{No. of data points} & \colhead{$\chi^{2}_{\nu}$\tablenotemark{d}}  \\
\colhead{\#} & \phn\phn\phn\phn$t$   & \phn\phn\phn\phn\phn$v$   &                            & $D$           &                 &               \\
             & \phn\phn\phn(d)   & \phn\phn\phn(\kms) &                            & (Mpc)         &                 &                        
}
\startdata
1 & \phn\phn7 -- 2,454 & 19,200 -- \phn8,800 & H$\alpha$ abs., em. (BVZI) & $3.96\pm0.05$  & 19 & 0.6   \\
2 & \phn\phn7 -- \phn\phn\phn19  & 19,200 -- 18,000   & H$\alpha$ abs.\phn (BVZI) & $3.92\pm0.06$  & \phn8 & 0.2    \\
3 & 523 -- 2,454& 11,100 -- \phn8,800    & H$\alpha$ em. \phn(BVZI)            & $4.01\pm0.07$ & 11 & 0.9 \\
4 & 523 -- 2,454& 10,800 -- \phn8,600    & H$\alpha$ em. \phn(RVZI)            & $3.91\pm0.07$ & 11 & 0.9 \\
5 & 553 -- 2,454& 12,000 -- \phn9,500    & H$\beta$  em. \phn(BVZI)            & $4.02\pm0.17$ & \phn7  & 1.6 \\
6 & \phn10 -- 2,454& 18,000 -- \phn8,600 & all of \# 1 - 5             & $3.95\pm0.04$\tablenotemark{e} & 37 & 0.9 \\
               &                &                                  &               &    &      &  \\
7 & 553 -- 2,069& 10,000 -- \phn7,500    & O[III] em. \phn(RVZI)               & $3.48\pm0.14$ & \phn6 & 1.2 \\
8 & 553 -- 2,454&  \phn9,200 -- \phn7,500    & Na I em. \phn(RVZI)                 & $3.48\pm0.11$ & 10  & 0.5 \\

\enddata
 
\tablenotetext{a}{Time and velocity ranges of the optical data used for the fit.}
\tablenotetext{b}{The optical lines from
Table~\ref{t1vopt} which were taken for the fit for specific distance
solutions.  The type of lines are H$\alpha$ in absorption (abs.) and
emission (em.) and H$\beta$ in emission (em.). The velocities for
absorption are BVZIs measured at the blue edge of the absorption
trough (Lewis \etal\ 1994) or the blue edge of the polarization
profile (Trammell \etal\ 1993).  The velocities from the emission
profiles are BVZIs measured at the blue edge of the profile (Matheson
\etal\ 2000a) and RVZIs measured at the red edge of the profile (Patat
\etal\ 1995).}
\tablenotetext{c}{The distance solutions from the weighted
least-squares fits of the radio angular velocities to the optical velocities. For the fits
the time ranges of the radio and optical data were chosen to be essentially equal. For the 
time range $7\leq t\leq 19$~d the radio angular velocities were extrapolated from 
later values (see text). The uncertainties are statistical standard
errors. Note: the errors are {\em not} scaled to $\chi^{2}_{\nu}$ (see
last column) of unity.}
\tablenotetext{d}{Chi-squared per degree of freedom.}
\tablenotetext{e}{Solution from all data used in the previous solutions 
combined.}

\end{deluxetable}

\begin{deluxetable}{c c c l}
\tablecaption{Distance error budget \label{t4err}}
\tablewidth{0pt}
\tablehead{ 
\colhead{\#} & \multicolumn{2}{c}{Error component}\tablenotemark{a} & \colhead{Error source} \\
             &      (Mpc)      &       (\%)       &                   
}
\startdata

1            & 0.05            & 1.3              & statistical\tablenotemark{b} \\
2            & 0.12            & 3.0              & anisotropic transverse expansion\tablenotemark{c} \\
3            & 0.05            & 1.3              & anisotropic radial expansion\tablenotemark{d} \\
4            & 0.05            & 1.3              & difference from BVZI abs. and em.\tablenotemark{e} \\
5            & 0.25            & 6.3              & radio-optical relation from boundary conditions\tablenotemark{f} \\
             &                 &                  &                                   \\
2 -- 5       & 0.29            & 7.2              & rss\tablenotemark{g} ~of systematic error contributions\tablenotemark{c} \\
1 -- 5       & 0.29            & 7.3              & rss of statistical and systematic error contributions                  \\

\enddata
\tablenotetext{a}{All error components are standard errors.}
\tablenotetext{b}{From solution 1 in Table~\ref{t3dfit}, largely includes error due to small-scale 
anisotropic transverse expansion (see item 2. in \S~\ref{systematic}).}
\tablenotetext{c}{{}Largely includes error due to large-scale anisotropic transverse 
expansion (see item 1. in \S~\ref{systematic}).}
\tablenotetext{d}{Difference between distance solutions with BVZI (solution 3 in Table~\ref{t3dfit}) 
and RVZI values (solution 4 in Table~\ref{t3dfit}). This contribution largely includes 
error due to large-scale anisotropic radial expansion (see item 3. in \S~\ref{systematic}).}
\tablenotetext{e}{The larger of the differences of the distance solutions with BVZI abs. and 
BVZI em. values to solution 1 (solution 3 minus solution 1 in Table~\ref{t3dfit}). 
This contribution includes systematic errors of the angular velocity fits (see item 5. in \S~\ref{systematic}). }  
\tablenotetext{f}{The uncertainty of relating optical and radio velocities (see item 7 in \S~\ref{systematic}).}
\tablenotetext{g}{Root-sum-square.}
\end{deluxetable}

\begin{deluxetable}{c l l l}
\tablecaption{Recent determinations of the distance to M81 \label{t5distdet}}
\tablewidth{0pt}
\tablehead{ 
\colhead{\#} & \colhead{Method} & \colhead{Distance (Mpc)} & \colhead{Reference} }

\startdata

1 & ESM             & 3.96$\pm$0.29  & this work \\
2 & HST Cepheids    & 3.63$\pm$0.34  & Freedman et al. (1994) \\
3 & HST Cepheids    & 3.93$\pm$0.26  & Huterer et al. (1995)\tablenotemark{a} \\
4 & PN luminosities\tablenotemark{b} & 3.5\phn$\pm$0.4  & Jacoby et al. (1989)  \\
5 & IR Tully-Fisher\tablenotemark{c}          & 3.7\phn$\pm$0.5 & Aaronson, Mould, \& Huchra (1980)  \\
6 & EPM\tablenotemark{d} & 2.6\phn$\pm$0.4     & Schmidt et al. (1993)\tablenotemark{e}  \\
7 & EPM                  & 4.2\phn$\pm$0.6     & Wheeler et al. (1993)\tablenotemark{f}  \\
8 & EPM                  & 3.65$\pm$1.45\tablenotemark{g}      & Prabhu et al. (1995)   \\
9 & EPM                  & 3.5\phn$\pm$0.2     & Clocchiatti et al. (1995)\tablenotemark{h}  \\
10 & SEAM\tablenotemark{i}                 & 4.0\phn$\pm$0.5     & Baron et al. (1995)  \\

\enddata
\tablenotetext{a}{Based on the Cepheid observations of Freedman et al. (1994).}
\tablenotetext{b}{Planetary nebula luminosities.}
\tablenotetext{c}{Infrared Tully-Fisher method.}
\tablenotetext{d}{Expanding photosphere method.}
\tablenotetext{e}{Eastman et al. (1996) noted that a peculiar Type II supernova like SN~1993J 
may not be a good candidate for either applying EPM or assessing its accuracy.}
\tablenotetext{f}{A note is given in the paper that the result depends on the detailed 
structure of the atmosphere.}
\tablenotetext{g}{The published value is 2.2-5.1 Mpc.}
\tablenotetext{h}{The complete equation for the distance is $D=\zeta(3.48 - 0.09\delta t_0
+0.09\delta R_0 \pm0.20)$~Mpc where the scattering parameter $\zeta\leq1$, and where $\delta t_0$ is the difference
between the actual time of shock breakout and JD=2,449,074.6 in tenths
of a day and $\delta R_0$ is the difference between the actual radius
of the progenitor star and 3.86$\times 10^{13}$~cm, in units of
$10^{13}$~cm.}
\tablenotetext{i}{Spectral-fitting expanding atmosphere method.}
\end{deluxetable}

\end{document}